\documentclass[prb,twocolumn,showpacs,preprintnumbers,amsmath,amssymb]{revtex4}
\usepackage{graphicx}
\usepackage{dcolumn}
\usepackage{bm}
\usepackage{bbold}
\usepackage{booktabs}

\begin{document}

\title{Nature of the low-energy excitations of \\
two-dimensional diluted Heisenberg quantum antiferromagnets}

\author{Ling Wang}
 \affiliation{Department of Physics, Boston University, 590 Commonwealth Avenue,
Boston, Massechussett, 02215}
\author{Anders W.~Sandvik}
 \affiliation{Department of Physics, Boston University, 590 Commonwealth Avenue,
Boston, Massechussett, 02215}

\date{\today}

\begin{abstract}
  We study the low-energy dynamics of $S=1/2$ antiferromagnetic Heisenberg clusters 
  constructed by diluting a square lattice at vacancy concentration $p$ at and 
  below the percolation threshold $p^* \approx 0.407$. The finite-size scaling 
  behavior of the average excitation gap, $\langle \Delta\rangle \sim L^{-z}$,
  where $L$ is the cluster length, is obtained using quantum Monte Carlo
  results for an upper bound $\Delta^*$ to $\Delta$, derived from
  sum rules. At the percolation threshold, we obtain a dynamic exponent
  $z=3.6 \pm 0.1 \approx 2D_f$ for clusters with singlet ($S=0$) ground state. 
  Here $D_f=91/48$ is the fractal dimensionality of the
  percolating cluster. We argue that this large dynamic exponent---roughly 
  twice that expected for quantum-rotor excitations---is a consequence of 
  weakly interacting localized effective magnetic moments, which form due to 
  local sublattice imbalance. This picture is supported by an extremal-value 
  analysis of local spectral gaps, which delivers an exponent relation 
  (between $z$ and two exponents characterizing the local gap distribution) 
  reproduced by our simulation data. However, the average $\langle\Delta^*\rangle$ 
  over all clusters, which have mostly ground state spin $S>0$, scales with a 
  smaller exponent than for the $S=0$ clusters alone;  $z\approx 1.5D_f$. 
  Lanczos exact diagonalization for small clusters show that typically, $S \to S-1$ 
  in the lowest-energy excitations, while the dominant spectral weight originates 
  from $S \to S+1$ excitations. Thus, the scaling of
  $\langle \Delta^*\rangle$ for clusters with ground state $S>0$ does  not reflect 
  the lowest-energy excitations, but the higher $S \to S+1$ excitations. This 
  result can be understood within a valence-bond picture. To further explore the scenario 
  of localized moments, we introduce a classical dimer-monomer aggregation 
  model to study the distribution of nearest-neighbor sites forming dimers (which are the objects 
  used in mapping to the quantum-rotor model) and unpaired spins (monomers). The monomers are 
  localized, and, thus, effective magnetic moments should form in the spin system. We  also 
  study the lowest triplet excitation of $S=0$ clusters using quantum Monte Carlo calculations
  in the valence bond basis. The triplet is concentrated at some of the classical monomer regions, 
  confirming the mechanism of moment formation. The number
  of spins (and moment regions) affected by the excitation scales as a non-trivial
  power of the cluster size. For a dimer-diluted bilayer Heisenberg model with weak 
  inter-layer coupling (where the system remains N\'eel ordered), there is no 
  sublattice imbalance. In this case we find $z \approx D_f$, consistent with 
  quantum rotor excitations. For a single layer at $p < p^*$ we find $z\approx 2 = D$, 
  which indicates that the weakly interacting localized moment mechanism is valid only 
  exactly at the percolation point. There is a cross-over behavior close to the 
  percolation point.
\end{abstract}

\pacs{75.40.Gb, 75.10.Jm, 75.10.Nr, 75.40.Mg}
                             
\maketitle

\section{\label{perco-intro}Introduction}

Two-dimensional (2D) antiferromagnets under doping with non-magnetic impurities 
have emerged as interesting systems with rich possibilities to explore various 
disorder-driven phase transitions belonging to different universality
classes.\cite{andersPRL2002,vajkPRL2002,vojtaPRB2006,andersPRL2006,vojtaPRL2005,yuPRL2005} 
Non-magnetic impurities (vacancies) enhance quantum fluctuation by reducing the connectivity of 
the spins. Many earlier calculations \cite{qtransition} for the 2D $S=1/2$ Heisenberg 
model had indicated that the quantum fluctuation can become strong enough to destroy 
the antiferromagnetic long-range order at a vacancy concentration $p_c$ less than the 
classical percolation threshold $p^*$---whence $p_c$ would be a quantum critical point.  
However, more recent quantum Monte Carlo (QMC) simulations of the diluted quantum Heisenberg 
model,\cite{kato00,andersPRB2002} studies of effective classical systems, 
\cite{vojtaPRB2006} as well as experiments on La$_{\rm 2}$Cu$_{1-x}$Zn$_{x}$O$_{\rm 4}$ 
(with non-magnetic Zn substituting $S=1/2$ Cu ions) \cite{vajkScience2002} all suggest 
that long range order actually survives all the way up to the percolation point $p^*$,
i.e., $p_c\equiv p^*$ for the single 2D layer.

The percolating cluster at $p^*$ is ordered,\cite{andersPRB2002} which implies that 
the static properties at the dilution-driven transition in the quantum Heisenberg model 
scale as in the classical (percolation) problem. However, quantum fluctuations lead to 
changes in the low-energy spin dynamics. The critical exponents therefore in general depend on 
classical percolation exponents as well as the dynamic exponent $z$ of the quantum spin 
clusters.\cite{vojtaPRL2005} The dynamic exponent of the percolating cluster is therefore 
important, and the focus of this paper.

The dynamic exponent governs the scaling of the gap $\Delta$ between the ground state
and the lowest excited state of a finite cluster. With $L$ denoting the cluster length 
(defined in some suitable way for a random cluster with irregular shape), the gap scales, 
on average, as $\langle \Delta \rangle\sim L^{-z}$. For a clean $D$-dimensional antiferromagnetic 
system on a bipartite lattice with $N$ (even) sites, every spin can be paired up with a 
nearest-neighbor spin on the opposite sublattice to effectively form a ``quantum rotor'' with 
angular momentum $l=0, 1$ states. In the mapping to a quantum rotor model,\cite{sachdevbook} 
these local degrees of freedom are replaced with angular momenta $l_i$ taking all integer values, 
with the high $l_i$ states suppressed due to their energy being $\propto l_i^2$. The ground state of 
the coupled quantum rotor system is a singlet. If the system is long-range ordered (but the global 
rotational symmetry has not been broken by any external perturbation), then the low-energy excitations 
of the coupled rotors (and the N\'eel ordered spin system\cite{anderson52}) are those of a 
single quantum rotor with mass $\propto N$. Thus $\Delta \sim N^{-1}$, i.e., $z=D$. 

According to one recently proposed scenario for randomly diluted antiferromagnets,\cite{vojtaPRL2005} 
the quantum rotor states remain the lowest-energy excitations even at $p^*$, where the dimensionality 
$D_f$ of the percolating cluster is fractal; $z=D_f=91/48$.\cite{stauffer} Following the discussion
above, this would seem to require that each spin can be paired up into a dimer with one of its nearest 
neighbors to effectively form a quantum rotor with $l=0$ ground state. This situation can be realized in 
the special case of the dimer-diluted bilayer,\cite{andersPRL2002} in which two coupled layers are 
diluted exactly in the same way by removing inter-layer spin dimers. All the remaining spins can then 
be paired with spins on the opposite layer. At sufficiently weak inter-layer coupling, the ground state 
of the largest connected cluster of spins in this system is long-range ordered for 
$p\leq p^*$,\cite{andersPRL2002} and, thus, the ground state should fall into the class of quantum rotor 
states with gap $\propto N^{-1}$. However, in the case of a single diluted layer (or a bilayer with
inter-layer coupling $J_\perp=0$), there are in general some ``dangling spins'' (or more generally, regions 
with local sublattice imbalance) in which not all spins can be simultaneously paired up into nearest-neighbor 
dimers. One may still be able to pair spins over longer distances (which would also imply longer-range 
interactions between the rotors in the effective model), but at some point, when very long distances are 
required, the mapping to simple quantum rotors should break down. 

Our assertion is that, at the percolation point, there are regions of spins that effectively form isolated 
magnetic moments, which cannot be described within an effective model containing only coupled rotors. The 
spatial distribution of these moment regions, and weak effective interactions between them (mediated by the 
magnetically inert parts of the percolating cluster), lead low-energy excitations which are dramatically 
different from those of the quantum rotor system. We introduced this scenario and presented supporting 
numerical evidence in a recent paper.\cite{lingPRL2006} Using finite-size scaling, we found a considerably 
larger dynamic exponent than the quantum-rotor value; $z \approx 2D_f$ instead of $z=D_f$. Here we provide 
more details of this work, and also expand significantly on the previous calculations. We use several 
different methods to indirectly and directly examine the low-energy excitations of different types of 
clusters, both at and away from the percolation point.

The conclusion that $z \approx 2D_f$ for clusters at the percolation point is based 
largely on quantum Monte Carlo (QMC) calculations of an upper-bound  $\Delta^*$ to the lowest 
excitation gap $\Delta$ for finite clusters with singlet ($S=0$) ground states. The bound is 
defined using standard sum rules, discussed in detail in Sec.~\ref{ssemethod} [and summarized 
as Eqs.~(\ref{gapbound}), (\ref{sumrule1}), and (\ref{sumrule2})]. The bound is exact,
$\Delta^*=\Delta$, for a spectrum with a single mode, and is known to scale with the
system size in the same way as $\Delta$ more generally, e.g., in the clean Heisenberg 
model.\cite{hasenfratz}  It can be evaluated for large clusters using QMC calculations,
in contrast to the exact gap, which is difficult to evaluate directly (because it is 
dominated by statistical errors if the gap is small). We also found that the probability 
distribution of local gaps $\Delta_i$ (also defined using a sum rule) scales with the 
system size.\cite{lingPRL2006} Defining $\epsilon_i = \Delta_i L^{a}$ (where the exponent
is determined from simulation data and is $a \approx 2.8$ for $S=0$ clusters), the distribution 
$P(\epsilon_i)$ is size-independent. Moreover, the low-energy tail of this distribution is well 
described by a power-law, $P(\epsilon_i)\propto {\epsilon_i}^{\omega}$, with $\omega =1$. 
Analyzing the local gaps using extremal-value statistics, we found that the dynamic 
exponent should be related to the parameters of the local gap distribution according to 
$z=a+D_f/(\omega +1)$. Our simulation results satisfy this exponent relation remarkably well. 
The applicability of the exponent relation supports the notion that the low-energy excitations 
involve a number $\propto n$ finite regions (containing the effective moments), while an exponent 
$a>0$ shows that individual excitations are not localized (since for localized excitations the
energy should be independent of $L$ for large $L$). The effective moments should be located in regions 
of imbalance in the number of spins on the two sublattices, and many moments can be involved in 
an excitation. The value of the exponent $a$ reflects the way in which the weak interactions between 
the effective moments involved in a particular excitation decrease with increasing system size, as 
these moments become further separated from each other.

In this paper, we report scaling results for larger clusters than previously and also compare results 
for clusters constructed in different ways. On the bipartite square lattice, we denote the number of 
sites on sublattice $A$ and $B$ as $n_A$ and $n_B$, respectively. The ground state has spin $S=|n_A-n_B|/2$. 
We analyze in detail both $S=0$ and $S>0$ clusters at the percolation point $p^*$. We use the gap 
upper-bound $\Delta^*$ from sum rules, as well as Lanczos exact diagonalization results for the 
excitation spectrum. For clusters with ground-state spin $S>0$, we point out that the spectral weight 
entering in the sum-rule approach is dominated by $S \to S+1$ excitations, whereas the lowest-energy 
excitations typically correspond to $S \to S-1$. The quantity $\Delta^*$ in this case describes only 
excitations where $S \to S+1$, for which we find $z \approx 1.5 D_f$ based on finite-size scaling. However, 
the lower $S \to S-1$ excitations most likely follow the same $z \approx 2D_f$ scaling as the $S=0 \to 1$
excitations of $n_A=n_B$ clusters. We also discuss results for the dimer-diluted bilayer at $p^*$, as well as 
the single layer at $p < p^*$. For these systems, we observe behavior consistent with quantum rotor excitations 
(although other scenarios, e.g., fractons,\cite{orbachModernPhy1994,terao94} are also possible). 

To explain the existence of localized moments in the percolating 
cluster, we also introduce a classical dimer-monomer aggregation model to study the purely geometrical 
local sublattice imbalance, which we believe is at the heart of this problem. The dimers correspond to 
nearest-neighbor sites that can form minimal local quantum rotors, and the monomers lead to "dangling" 
spins that are, due to local sublattice imbalance, left over after the maximum number of dimers has formed. 
The monomers, individual ones or groups  of several of them, can lead to effective magnetic moments in the 
spin system. We find that the classical monomers indeed are confined within regions of finite size, both 
at and away from the percolation point. The anomalous dynamics with $z \approx 2D_f$ in the single-layer
quantum spin system at $p^*$ should therefore be a consequence of localized quasi-free magnetic moments 
interacting very weakly because of the vanishing spin stiffness of the percolating cluster.\cite{andersPRB2002} 
Away from the percolation point, the moments can lock to the global N\'eel order of the cluster (as a
single magnetic impurity in two dimensional is known to do \cite{sachdev99,hoglund04}) and do not form an 
effective independent low-energy system.

To further investigate the nature of the excitations of the quantum spins and their relationship to the 
classical monomers, we have also applied a projector QMC method in the valence bond basis~\cite{andersvbmc1} 
to directly study the triplet excitations of clusters with singlet ground states. In the valence bond basis, 
a triplet state can be described by a lone triplet bond, the location of which fluctuates among the background 
singlet bonds. We find that the triplet bond is indeed predominantly localized at a subset of the classical 
monomer regions. The total size of the excitation (i.e., the number of spins involved in it) is not finite, however, 
but grows with the cluster size according to a non-trivial power law.

The outline of the rest of the paper is as follows. After defining the spin models and describing 
several computational methods in Sec.~\ref{perco-model}, we present results of both Lanczos exact 
diagonalization and sum-rule QMC calculations for single-layer clusters at $p=p^*$ in Sec.~\ref{moredata}. 
In Sec.~\ref{spectralsection} we discuss the distribution of spectral weight in the dynamic structure
factor originating from excited states of different total spin, using Lanczos exact diagonalization as
well as an approximate analysis based on valence bond states. We discuss scaling results for percolating 
bilayer clusters in Sec.~\ref{bilayerp}, and for single-layer clusters away from the percolation point in 
Sec.~\ref{belowpc}. The classical dimer-monomer aggregation model is discussed in Sec.~\ref{classicalmodel}, 
and results of the valence-bond projector QMC simulations of triplet excitations in Sec.~\ref{rvbqmc}. We
conclude in Sec.~\ref{perco-summary} with a summary and discussion. 

\section{\label{perco-model}Model and methods}

The Heisenberg Hamiltonian on a single site-diluted layer is given by
\begin{equation}
  \label{perco-hamiltonian}
  H=J\sum_{\langle i,j\rangle}\delta_i \delta_j 
  {\bf S}_i\cdot {\bf S}_j , \quad (J>0),
\end{equation}
where $\langle i , j\rangle $ denotes nearest neighbors on a 2D square lattice and $\delta_i=0$ 
(vacancy) and $\delta_i=1$ (magnetic site) with probability $p$ and $1-p$, respectively. We study 
clusters with two types of boundary conditions. In open-boundary $L\times L$ systems, we start with 
all magnetic sites and introduce vacancies with probability $p$. We study the largest cluster of 
connected magnetic sites. The number of spins $n$ in such clusters fluctuates and scales as 
$\langle n\rangle \sim L^{D_f}$, with $D_f=91/48$. We also study clusters grown on an infinite lattice. 
Starting from a single magnetic site, we add more sites to the cluster with probability $1-p$ by 
transversing along the boundary sites, leaving sites unfilled with probability $p$, but flagging each 
site as visited (so that sites assigned as vacancies are not visited again). This 
procedure terminates at random at some stage where all neighbors of the cluster have been assigned as 
vacancies. We only keep clusters of some desired target size $n$. These clusters have a characteristic 
average length $\langle L\rangle$ (defined, e.g., as their radius of gyration) such that
$n  \propto \langle L\rangle ^{D_f}$. The two types of clusters will be referred to as 
$L\times L$ and fixed-$n$, respectively. In Ref.~\onlinecite{lingPRL2006}, we only studied fixed-$n$ 
clusters. Here we also consider the $L\times L$ variant to check whether the finite-size scaling 
properties depend on the boundary conditions in the cluster construction. For $p<p^*$,
we consider only the $L\times L$ clusters, because the fixed-$n$ construction rarely terminates at
reasonably small $n$ in this case.

Under each type of boundary condition, we further consider two different ensembles of sublattice 
occupations; $n_A=n_B$, in which case all clusters have ground state spin $S=0$, as well as arbitrary $n=n_A+n_B$
(with the distribution give by the cluster construction), corresponding to ground state spin $S=|n_A-n_B|/2$. 
The latter ensemble includes also the $S=0$ clusters.

A bilayer cluster is constructed by coupling two identical single-layer clusters 
with an interlayer coupling constant $J_{\perp}$. The Hamiltonian is thus
\begin{eqnarray}
\label{bilayerh}
\nonumber
  H&=&J\sum_{\langle i,j\rangle}\delta_i \delta_j \big(
{\bf S}_{1i}\cdot {\bf S}_{1j}+{\bf S}_{2i}\cdot {\bf S}_{2j}\big)\\
&&+J_{\perp}\sum_i\delta_i{\bf S}_{1i}\cdot {\bf S}_{2i},
\label{hambilayer}
\end{eqnarray}
where the subscripts $1,2$ refer to the two layers. Also in this case we can study 
$L\times L$ or fixed-$n$ clusters, but, in contrast to the single layer, the ground 
state of a bilayer cluster is always a singlet because each spin can be paired with 
its neighbor in the opposite layer. We consider small coupling ratios $J_{\perp}/J$,
for which the ground state has long-range order.\cite{andersPRL2002}

Here our main interest is in the the energy gap $\Delta$ between the ground state and the 
first excited state, which in the case of an $n_A=n_B$ cluster is a singlet-triplet gap. For
clusters with general $n_A,n_B$ such that $S=|n_A-n_B|/2>0$, the lowest excitation can have 
total spin $S^{\prime}=S-1,S$, or $S+1$. In addition to the gap, the distribution of the spin 
$S'$ of the lowest-energy excitation is also interesting. We will also study the localization
properties of the excitations very explicitly, by formulating the problem in the valence bond 
basis and carrying out unbiased quantum Monte Carlo calculations of $S'=1$ excitations
of clusters with $S=0$ ground states.

To calculate the gaps, we use both direct and indirect (approximate, through sum-rules) estimates, 
using  the methods discussed in Secs.~\ref{ed} and \ref{ssemethod}. In Sec.~\ref{ipr} we will introduce 
the valence bond QMC scheme for directly imaging the spatial distribution of triplet excitations.

\subsection{\label{ed}Exact diagonalization}

The most straight-forward approach is to diagonalize the Hamiltonian numerically in sectors 
of different magnetization, 
\begin{equation}
m_z=\sum_{i=1}^n S^z_i,
\end{equation}
using the Lanczos method. However, for irregular
clusters (without lattice symmetries to exploit for block-diagonalization), this can be 
done in practice only for up to $n \approx 20$ spins, due to the rapid growth of the matrix 
sizes with $n$ (considering also that we have to average over a large number---typically 
thousands---of random cluster realizations). Nevertheless, such calculations are very useful and
give some important insights into the role of ``dangling'' spins in low-energy excitations.  

In addition to studying the level spectrum, focusing on a few low-lying states and calculating 
their total spin to classify the excitations, we also compute the full dynamic spin structure 
factor (in the standard way with the Lanczos method, as described, e.g., in 
Ref.~\onlinecite{dagottoreview});
\begin{equation}
S({\bf q},\omega) = \sum_{m} |\langle m|S^z_{\bf q} |0\rangle|^2 \delta(\omega+E_0-E_m),
\label{dynamicsS}
\end{equation}
where $S^z_{\bf q}$ is the Fourier transform of the spin operators; 
\begin{equation}
S^z_{\bf q} = \frac{1}{\sqrt{n}}\sum_{j=1}^n {\rm e}^{i{\bf q}  \cdot {\bf r}_j} S^z_j.
\label{sqop}
\end{equation} 
In a clean Heisenberg antiferromagnet on a bipartite lattice, the lowest excitation is a triplet 
at $q=(\pi,\pi)$. We can use this wave-vector also for the diluted system, although the momentum 
is no longer conserved, i.e., the energy eigenstates $|m\rangle$ in (\ref{dynamicsS}) are not 
classified by the quantum number ${\bf q}$, but the spin operators $S^z_{\bf q}$ are still completely well 
defined. We expect $S(\pi,\pi,\omega)$ to exhibit the largest spectral weight for the low-energy 
excitations, since these should involve out-of-phase fluctuations of neighboring spins. As 
we will see in Sec.~\ref{generals}, the dynamic structure factor is of great utility in judging the 
validity of our sum-rule based approach for an upper-bound of the energy gap, which we discuss next.

\subsection{\label{ssemethod}Quantum Monte Carlo and sum rules}

We use the stochastic series expansion (SSE) QMC method \cite{andersPRB1999} to calculate quantities 
which are closely related to the gap.  An upper-bound $\Delta^*$ to the ground state energy gap $\Delta$ 
can be obtained using the static spin structure factor $S({\bf q})$ and susceptibility $\chi(\bf q)$ at 
the staggered  wave-vector ${\bf q}=(\pi,\pi)$;
\begin{equation}
\Delta^* = 2S(\pi,\pi)/\chi(\pi,\pi) \ge \Delta.
\label{gapbound}
\end{equation}
This bound follows from the well-known sum-rules;
\begin{eqnarray}  
  \int_0^{\infty}d\omega S({\bf q},\omega)&=&S({\bf q}), \label{sumrule1} \\
  2\int_0^{\infty}\frac{d\omega}{\omega}S({\bf q},\omega)&=&\chi({\bf q}),
  \label{sumrule2}
\end{eqnarray}
which, in the way written here, are valid at temperature $T=0$. In a system with a sole triplet 
mode (a hypothetical situation) with energy $\omega_{\bf q}$, 
we get $2S({\bf q})/\chi({\bf q})=\omega_{\bf q}$. 
Any spectral weight above this lowest mode will render the ratio larger than $\omega_{\bf q}$. 
For a clean system, the lowest quantum rotor state is at ${\bf q}=(\pi,\pi)$ (whereas at other
wave-vectors spin-waves are the lowest excitations). As we discussed above, we expect 
${\bf q}=(\pi,\pi)$ to be the best choice for examining low-energy excitations also in the 
diluted system, and we here focus exclusively on this case.

The staggered structure factor and susceptibility can be efficiently calculated with the SSE
method with "operator-loop" updates.\cite{andersPRB1999} Using the standard definitions, for a given 
cluster of $n$ sites the static staggered structure factor is
\begin{equation}
\label{sfactor}
S(\pi,\pi)=\frac{1}{n}\left \langle \left(\sum_{i=1}^n(-1)^{\phi_i}S_i^{z}\right)^2\right \rangle,
\end{equation}
and the corresponding susceptibility is given by
\begin{equation}
\label{staggerchi}
\chi(\pi,\pi)=\frac{1}{n}\Big \langle \sum_{i,j=1}^n(-1)^{\phi_{j}-\phi_i}
\int_0^{\beta}d\tau S_i^z(\tau)S_j^z(0) \Big \rangle,
\end{equation}
where $\phi_{i}=x_i+y_i$. Disorder averages are subsequently calculated for the ratio in 
(\ref{gapbound}) (where, it should be stressed, we first evaluate the ratio separately for each 
cluster, in order to obtain the gap bound specifically for each of them, and then take the 
average) using, typically, thousands of random realizations of either the largest cluster 
on $L\times L$ lattices or fixed-$n$ clusters. 

For a clean Heisenberg antiferromagnet, $\Delta^*$ is known\cite{hasenfratz} to scale with the 
system size as the true gap; $\langle \Delta^*\rangle \sim \langle\Delta\rangle \sim L^{-z}$. This is 
because the dominant spectral weight is at the very lowest excitation energy---the spectral function 
in the thermodynamic limit has a delta-function at the lowest energy, followed by a continuum at higher 
energies. We expect similar spectral features in the percolating cluster and suspect that $\Delta^*$ 
should scale as $\Delta$ (and will show supporting numerical results in the next section).  At the very 
least, {\it if the true power-law behavior is $\langle \Delta\rangle \sim L^{-z} \sim n^{-z/D_f}$, then the 
value $\tilde z$ extracted from finite-size scaling of $\Delta^*$ must be a lower bound to the true dynamic 
exponent}. Actually, in Sec.~\ref{moredata} we will use Lanczos results for the dynamic structure factor 
on small clusters to show that the finite-size scaling of $\Delta^*$ {\it does not} reflect the true 
lowest-energy excitations in the case of $S>0$ clusters, but all indications are that the sum-rule 
approach is valid for $S=0$ clusters.

We will also study an effective local (site-dependent) excitation gap
\begin{equation}
\Delta_i = \frac{1}{2}\frac{1}{\chi_i},
\end{equation}
which is analogous to the gap bound (\ref{gapbound}) but here the
"local structure factor" is just a constant; $S_i =
(S^z_i)^2=1/4$. The local susceptibility $\chi_i$ is defined as
\begin{equation}
\chi_i=\int_0^{\beta}d\tau\langle S_i^z(\tau)S_i^z(0)\rangle.
\label{localx}
\end{equation}
Although the imaginary-time dependent correlation function $\langle S_i^z(\tau)S_i^z(0)\rangle$ is asymptotically, for
$\tau \to \infty$, dominated by the lowest excitation, in practice the integral will be dominated by the excitation(s) 
which predominantly affects the given site $i$. For a disordered system, different sites can be affected by different 
excitations, and $\Delta_i$ then represents a typical energy scale of excitations affecting spin $i$.

We should note that for clusters with ground state spin $S>0$, the grand-canonical SSE method samples 
over all magnetization sectors $-S \le m_z \le S$.\cite{andersPRB1999} We therefore have to subtract 
the static $(\omega=0)$ contributions in Eqs.~(\ref{sfactor}), (\ref{staggerchi}), and (\ref{localx}) 
arising from a non-zero $m_z$, i.e., in Eq.~(\ref{localx}) we subtract $\langle S^z_i\rangle^2$ computed
in the different $m_z$ sectors and averaged over all $m_z$.

The SSE method operates at $T>0$, but we can achieve the $T \to 0$ limit by choosing $T$ sufficiently low 
for all quantities of interest to converge. We use a "$\beta$ doubling" procedure,\cite{andersPRB2002} 
in which the inverse temperature is successively doubled until there is no longer any detectable 
dependence of calculated quantities on $\beta$. Since the dynamic exponent 
is large, the temperature $T \ll \Delta$ has to be very low indeed for large clusters. As an 
example of the ultra-low temperatures required, the largest $\beta$ we use for $n=512$ clusters 
with $S=0$ is $\beta=2^{19} \approx 5 \times 10^5$. Since the simulation (CPU) time and memory 
usage scale essentially linearly in both $\beta$ and $n$, these calculations are quite demanding. 
Fortunately, the SSE code for the isotropic Heisenberg model can be effectively parallelized,\cite{lingthesis} 
and we have run most of the simulations on a massively parallel computer very well suited for 
these calculations.\cite{bluegene}

When studying disorder averaged static properties, the SSE runs for each individual cluster can be rather short. 
As long as each run is properly equilibrated (for which the $\beta$-doubling procedure also helps \cite{andersPRB2002}), 
the average over many realizations will give an unbiased estimate to any simple average, e.g., 
a spin correlation function. However, when computing nonlinear functions involving several quantities, such as 
the ratio (\ref{gapbound}), the statistical errors introduce a bias. It is therefore important to collect sufficient 
statistics for the individual clusters. We have compared results of runs of different lengths in order to make sure 
that the results presented here do not suffer from significant bias effects.

\subsection{\label{ipr}Valence-bond projector Monte Carlo}

To study the nature of the lowest triplet excitation of clusters with $S=0$ ground states, 
we apply a valence bond projector Monte Carlo method.\cite{liang1,andersvbmc1} This method has 
been described in detail in recent papers \cite{andersvbmc2,andersvbmc3} and we here only review 
the elements necessary to understand the way we can access the triplet excitations and study 
their spatial distribution on the clusters.

First, consider the singlet ground state $|0\rangle_s$, which we want
to project out from a singlet ``trial'' state $|\Psi\rangle_s$. The latter has
an expansion in all singlet energy eigenstates;
\begin{equation}
|\Psi\rangle_s = \sum_n c_n |n\rangle_s.
\end{equation}
In the standard way,
if the ground state energy is the eigenvalue of the Hamiltonian which is the largest 
in magnitude, which can always be assured by subtracting a constant from $H$ 
(which we assume has been done, if necessary), the ground state can be projected 
out by applying a high power of $H$ to the trial state;
\begin{eqnarray}
&&(-H)^P |\Psi\rangle =  c_0(-E_0)^P \times \label{projection1} \\ 
&&~~~~~~\left [ |0\rangle_s + \frac{c_1}{c_0}\left (\frac{E_1}{E_0} \right )^P 
|1\rangle_s +\ldots \right ],\nonumber
\label{projhp}
\end{eqnarray}
where we include the minus sign because normally $E_0<0$. For large $P$ all the excited states 
are filtered out because the ratios $|E_n/E_0|<1$.

Valence-bond basis states are products of $N/2$ singlets,
\begin{equation}
(i,j)=(\uparrow_i\downarrow_j-\downarrow_i\uparrow_j)/\sqrt{2},
\end{equation}
where we consider the first, $i$, and second, $j$, spins to always be on sublattice $A$ and $B$, 
respectively. The trial state is thus expressed in this over-complete basis as
\begin{eqnarray}
|\Psi\rangle_s & =  & \sum_v w_v |(i^v_1,j^v_1),\ldots,(i^v_{N/2},j^v_{N/2})\rangle \nonumber \\
               &  = & \sum_v w_v |V_v\rangle, \label{psisinglet}
\end{eqnarray}
where $v$ labels all the different tilings of the cluster into valence bonds, of which there 
are $(N/2)!$, and we have introduced the short-hand notation $|V_v\rangle$ for a valence-bond 
basis state. 

For a trial state in a ground state projector calculation, it is convenient to use 
an amplitude product state,\cite{liang2,jievar} where the expansion coefficients are given by
\begin{equation}
w_v=\prod_{x,y}h(x,y)^{n_v(x,y)},
\label{wamp}
\end{equation}
where $h(x,y)>0$ and $n_v(x,y)$ is the number of bonds of size $(x,y)$ in the configuration, i.e., 
the length of the bond is $r=(x^2+y^2)^{1/2}$. Note that it is not necessary to normalize the
trial state. 

For the clean 2D system, the optimal amplitudes are translationally invariant and decay as $h(r) \sim r^{-3}$.\cite{jievar} 
For random clusters, the optimal amplitudes are naturally not translationally invariant. While the average bond probabilities 
(which are related to the amplitudes) decay with $r$, for any given cluster there are typically some regions spanned by long bonds 
(a feature intimately connected with the low-energy physics, as we will discuss in Secs.~\ref{spectralsection} and \ref{rvbqmc}). 
One could in principle optimize all the $\propto n^2$ different amplitudes for each specific cluster.  However, the effort 
involved in individual optimizations for hundreds or thousands of clusters does not necessarily pay off, compared to just 
projecting the trial state with a somewhat larger power $P$ of $H$. We here use a very simple trial state with all $h(x,y)=1$.

To carry out the projection using Monte Carlo sampling, we write the $S=1/2$ Heisenberg 
Hamiltonian in terms of singlet projection operators on all the pairs $b$ of nearest-neighbor 
sites $\langle i(b),j(b)\rangle$;
\begin{equation}
H_b \equiv H_{i(b),j(b)}=-\left ({\bf S}_{i(b)}\cdot {\bf S}_{j(b)}-\frac{1}{4} \right ),
\end{equation}
and write the projection operator in (\ref{projhp}) as
\begin{equation}
(-H)^P = \left (\sum_{b=1}^{N_b} H_{b} \right)^p 
= \sum_r {\cal P}_r,
\end{equation}
where 
\begin{equation}
{\cal P}_r = H_{b^r_{m}}\cdots H_{b^r_{2}}H_{b^r_{1}}
\end{equation}
denotes the possible strings, $r=1,\ldots,N_b^P$, of the singlet projectors. 

When a singlet projector $H_{ij}$ acts on a state with a valence bond on the two sites $i,j$, 
the state remains unchanged with a matrix element of unity; we call this a diagonal projection. 
If the operator acts on a state with no valence bond on the two sites, then the two bonds  $(i,k)$ 
and $(l,j)$ connected to $i,j$ are broken, and new singlets $(i,j)$ and $(l,k)$ are formed. 
This process has matrix element $1/2$, and we call it an off-diagonal projection. Thus the 
projection rules are;
\begin{eqnarray}
& & H_{ij}|...(i,j)... \rangle =|...(i,j)... \rangle, \label{hijdia} \\
& & H_{ij}|...(i,k)...(l,j)...\rangle =
\hbox{$\frac{1}{2}$}|...(i,j)...(l,k)...\rangle. \label{hijoff}
\end{eqnarray}
Acting on a component $|V_a\rangle$ of the trial state, a string ${\cal P}_r$ effects a number 
of rearrangements (\ref{hijoff}) of pairs of valence bonds, resulting in another valence bond 
basis state which we call $|V_a(r)\rangle$; 
\begin{equation}
{\cal  P}_r |V_a\rangle = W_{ar}|V_a(r)\rangle.
\label{wrk1}
\end{equation}
Here the ``projection weight'' $W_{ar}$ for a combination of operator string ${\cal P}_r$ and state 
$|V_a\rangle$ is given by the number $m_{\rm{off}}(a,r)$ of off-diagonal operations (\ref{hijoff}) 
in the course of the projection;
\begin{equation}
W_{ar}=2^{-m_{\rm off}(a,r)}.
\label{wrk2}
\end{equation}
The expectation value of an operator $A$ can be written
\begin{eqnarray}
\langle A\rangle & = & 
\frac{\sum_{ab}\sum_{rl}w_aw_b \langle V_b | {\cal  P}_l^* A {\cal P}_r |V_a\rangle}
{\sum_{ab}\sum_{rl} w_aw_b \langle V_b | {\cal  P}^*_l {\cal  P}_r |V_a\rangle} \label{vbaexpvalue}\\
&= & \frac{\sum_{ab}\sum_{rl} w_aw_b W_{ar}W_{bl}\langle V_b(l) | A |V_a(r)\rangle}
{\sum_{ab}\sum_{rl} w_aw_bW_{ar}W_{bl} \langle V_b(l)  |V_a(r)\rangle}. \nonumber 
\end{eqnarray}
where $w_a$ and $w_b$ are the weights computed according to (\ref{wamp}) for the bonds in the states $|V_a\rangle$ 
and $\langle V_b|$ in the expansion (\ref{psisinglet}) of the trial ket $|\Psi\rangle_s$ and bra $_s\langle\Psi|$ 
states. 

The sampling weight to be used in Monte Carlo calculations of (\ref{vbaexpvalue}) is
\begin{equation}
W(a,b,r,l)=w_aw_bW_{ar}W_{bl} \langle V_b(l)  |V_a(r)\rangle,
\label{vboverlap}
\end{equation}
where the overlap of the two projected states is given by
\begin{equation}
\langle V_b(l)  |V_a(r)\rangle = 2^{N_\circ - N/2},
\label{vboverlapvalue}
\end{equation}
where $N_\circ$ is the number of loops formed when the  bond configurations of the states
$|V_a(r)\rangle$ and $\langle V_b(l)|$ are superimposed (forming the transposition graph
\cite{liang2}). Simple sampling procedures for the operator strings and trial state bonds are 
described in Refs.~\onlinecite{andersvbmc1,andersvbmc2}. More efficient sampling methods have 
been developed recently,\cite{andersvbmc3} which we use but do not discuss here. 

For the purpose of the present paper, the most interesting aspect of the valence bond projector 
scheme is the fact that we can easily extend the scheme to also study a triplet state. A trial 
wave function in the triplet sector can be expressed in the overcomplete basis of a lone triplet 
bond among $N/2-1$ singlets. We denote a zero-magnetization triplet by square brackets;
\begin{equation}
[i,j]=(\uparrow_i\downarrow_j+\downarrow_i\uparrow_j)/\sqrt{2},
\end{equation}
and expand the triplet trial state as
\begin{eqnarray}
|\Psi\rangle_t &  = & \sum_v w_v \sum_{m=1}^{N/2}|(i^v_1,j^v_1)\ldots
[i^v_m,j^v_m]\ldots(i^v_{N/2},j^v_{N/2})\rangle \nonumber \\
               &  = & \sum_v w_v \sum_{m=1}^{N/2} |V_m\rangle,
\label{psitriplet}
\end{eqnarray}
where the normalization is again irrelevant. Here we use the same expansion coefficients---the 
amplitude products (\ref{wamp})---as in the singlet trial state. Note that for a clean system, 
the singlet state (\ref{psisinglet}) has momentum ${\bf k}=(0,0)$, whereas the triplet 
(\ref{psitriplet}) has ${\bf k}=(\pi,\pi)$. These are the known momenta of the lowest states in the 
two spin sectors (with the triplet being the lowest member of Anderson's tower of quantum 
rotor states\cite{anderson52}). The wave-function signs corresponding to (\ref{psisinglet}) and 
(\ref{psitriplet}) should be correct for the lowest singlet and triplet states also for a diluted 
system, since all conditions for Marshall's sign rule [which corresponds to all positive expansion
coefficients in Eq~(\ref{psitriplet})] \cite{marshall} remain valid.

When acting on a triplet bond, the singlet projector $H_{ij}$ destroys the state, while the action 
between a singlet and a triplet bond are very similar to the pure singlet rules (\ref{hijoff}).
The two triplet rules are
\begin{eqnarray}
& & H_{ij}|...[i,j]... \rangle =0, \label{thijdia} \\
& & H_{ij}|...[i,k]...(l,j)...\rangle =
\hbox{$\frac{1}{2}$}|...(i,j)...[l,k]...\rangle. \label{thijoff}
\end{eqnarray}
In the projector method, it is straight forward to convert one bond of the singlet trial 
wavefunction into a triplet and trace its evolution. The triplet states that survive after all 
$P$ operations [i.e., that are not destroyed by a diagonal operation (\ref{thijdia})] are used 
to measure properties in the triplet sector. To measure triplet expectation values, we have to 
project triplets like this both in the bra and ket in the triplet version of (\ref{vbaexpvalue}).
We also have to check the overlap (\ref{vboverlap}) of the surviving triplet states. One can 
show that the two triplet bonds have to be in the same transposition-graph loop in order for the overlap 
to be non-zero, and it is then equal to the singlet overlap (\ref{vboverlapvalue}). For surviving pairs 
of triplets, the weight of the triplet configuration is the same as that of the original singlet 
one. One can therefore sample the configurations in the singlet sector, and carry out measurements 
with all the surviving triplets {\it without reweighting}. This is one of the strengths 
of the valence bond projector method. 

There can still be problems with this approach, because the number of surviving triplets decreases 
with the projection power $P$ [because the probability of a triplet to be destroyed by a diagonal
triplet operation (\ref{thijdia}) increases). It helps considerably that the starting trial state can have 
the triplet at $N/2$ different locations, in both the bra and the ket state, and as long as one pair out 
of the total of $(N/2)^2$ combinations survives (and gives non-zero overlap), we can collect statistics. 
One can carry out the summation over triplet locations $m$ in (\ref{psitriplet}) efficiently, without introducing 
any additional factor $N/2$ in the computational effort, in a single traversal of the operator sequence. 

In some cases it can still happen that the triplet quantities of interest have not converged well to the $P\to \infty$ 
limit before the triplet survival probability becomes too low to be useful. This is not a serious problem in the 
present application, although an extrapolation to infinite $P$ based on several calculations with reasonable triplet 
survival probability is necessary to ensure that the results represent the lowest triplet. An exponential 
asymptotic convergence can be expected based on Eq.~(\ref{projhp}).

\begin{figure}
\begin{center}
\includegraphics[width=6.75cm]{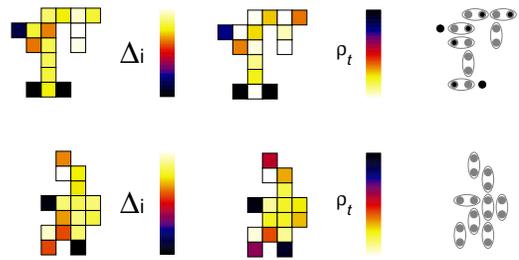}
\caption{(Color online) Results for  two different clusters, visualized with color scales, of the QMC sum rule approximation 
of local gaps $\Delta_i$ (left) and valence bond projection calculation of the triplet density $\rho_i$ (center). To the right,
the clusters are shown covered with dimers (pairs of spins enclosed by ovals) and left-over monomers (black circles). The black
circles inside ovals indicate other possible locations of the monomers, corresponding to alternative maximal dimer coverings (which 
here always corresponds to two left-over monomers).}
\label{clusterdemo}
\end{center}
\vskip-3mm
\end{figure}

We will discuss the spatial distribution of the triplets. The surviving triplet states have the triplet bond located at 
two particular sites (which can be different in the ket and the bra, and we do the measurements in both of these states). 
In a random system, the average triplet density will not be uniform and provides a very concrete measure of the localization properties 
of the lowest triplet excitation.

Note that the distribution of the $m_z=0$ triplet bond is equivalent to the magnetization distribution in a state with $m_z=1$, which 
could also be studied using the SSE method at low temperatures (e.g., by including a weak magnetic field \cite{syljuasenPRE2002}). 
However, the valence bond states also contain other relevant information, e.g., the statistics of the length of the triplet bond, 
which can only be accessed in the valence bond basis and which will be useful for analyzing the nature of the excitations (as we
will do in Sec.~\ref{rvbqmc}).

\subsection{Examples\label{sec_examples}}

Having introduced the technical aspects of all the methods, we now present illustrative results for two 
small clusters. This will help to clarify the subsequent analysis and discussion of results for 
larger clusters.

The local gaps $\Delta_i$ and the triplet density $\rho_i$ are visualized for two different clusters in 
Fig.~\ref{clusterdemo}. Here the color scales were created separately for the two clusters, with the minimum
and maximum values for each quantity on a particular cluster corresponding to the extrema of the scales shown 
(and, thus, the plots should only be used to examine the variations within the clusters, not comparing the values 
for the two clusters). 

The two clusters differ qualitatively in a way which is directly related to our arguments pertaining to a low energy 
scale. The lower cluster can be completely subdivided into pairs of nearest-neighbor sites (dimers, represented by ovals), 
whereas the upper one has two ``dangling spins'' left (monomers, shown as black circles outside ovals) after the sites 
have been paired up as much as possible. The pairing into dimers is not unique---the black circles inside ovals show all 
other possible monomer locations for this cluster. In all cases there are two monomers in two separate 
regions. The classical dimer-monomer aggregation model discussed in Sec.~\ref{classicalmodel} contains the statistics 
of the distribution of the monomers. Our main argument is that the presence of monomer sites leads to small gaps, i.e., 
a large dynamic exponent. For the two clusters shown, the exact gaps are $0.039J$  and $0.276J$, respectively, for the 
cluster with and without monomer sites. The gap upper bounds $\Delta^*$ are $0.076J$ and $0.35J$. While in particular 
the former is quite far from the exact result (in a relative measure), the difference between the two clusters 
is still large.

Large clusters are likely to have dangling spins, and the top cluster in Fig.~\ref{clusterdemo}
is therefore the more interesting case. One can clearly see a strong correspondence between
small local gaps and large triplet density, and they both coincide very well with sites where
monomers can be located. Although this is in accord with the notion of monomers leading to finite 
regions of spins affected by the excitation, these clusters are clearly too small to give any 
meaningful quantitative insights into the localization properties of the triplet.

In the following four sections we will carry out quantitative scaling analyses of the gaps in different
types of clusters, while further discussion of the monomer and triplet distributions will be postponed to 
Secs.~\ref{classicalmodel} and \ref{rvbqmc}, respectively.

\section{\label{moredata}Single-layer gap scaling at the percolation point}

We here discuss the distribution of exact gaps obtained with the Lanczos method, as well as SSE
QMC results for the gap upper bound and local gaps. First we consider $S=0$ clusters ($n_A=n_B$), 
and then arbitrary $S$.

\begin{figure}
\begin{center}
\includegraphics[width=5.5cm]{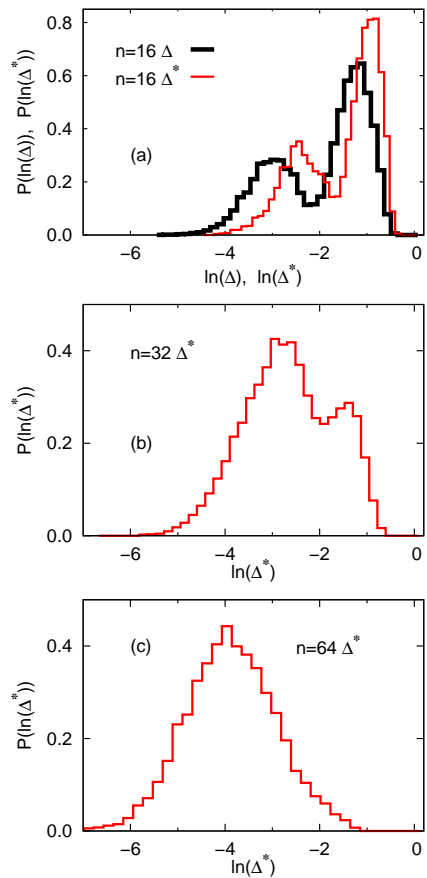}
\caption{(Color online) Distribution of the singlet-triplet gap $\Delta$ and its upper bound
QMC estimate $\Delta^*$ for $n=16$ clusters (a), and the $\Delta^*$ distribution for $n=32$ (b)
and $n=64$ (c).}
\label{lanczos-sse}
\end{center}
\vskip-3mm
\end{figure}

\subsection{\label{moredata_a}Clusters with singlet ground state}

Fig.~\ref{lanczos-sse}(a) shows the probability distribution of the logarithm of the exact gap 
$\Delta$ of $n=16$ clusters obtained using $4 \times 10^4$ samples. We also show results for the 
upper-bound $\Delta^*$ for clusters of the same size, obtained from SSE calculations for $6 \times 10^3$ 
different clusters. We presented these results in Ref.~\onlinecite{lingPRL2006} and here re-graph them in 
a different way for added clarity. The $\Delta^*$ curve is visibly shifted up in energy relative 
to the $\Delta$ distribution (with the average $\Delta^*/\Delta \approx 1.5$), but the shapes of the 
two curves are remarkably similar. The two-peak structure is related to the ``dangling spins'' 
discussed in Sec.~\ref{sec_examples}. The large-gap peak originates almost exclusively from clusters
that can be completely partitioned into nearest-neighbor dimers, whereas the low-gap peak corresponds 
to clusters with dangling spins (monomers). Clearly, as the cluster size grows, it will be less and 
less likely to find clusters with no monomers, and the weight of the high-energy peak should therefore 
gradually diminish and be absent for large clusters. The relative size of the large-$\Delta^*$ peak is 
indeed much smaller in the $n=32$ distribution graphed in Fig.~\ref{lanczos-sse}(b). In the $L=64$ 
histogram, shown in panel (c), only a single peak can be discerned (with only a weak tail suggesting some 
remaining contributions from no-monomer clusters). 

\begin{figure}
\begin{center}
\includegraphics[width=6.5cm]{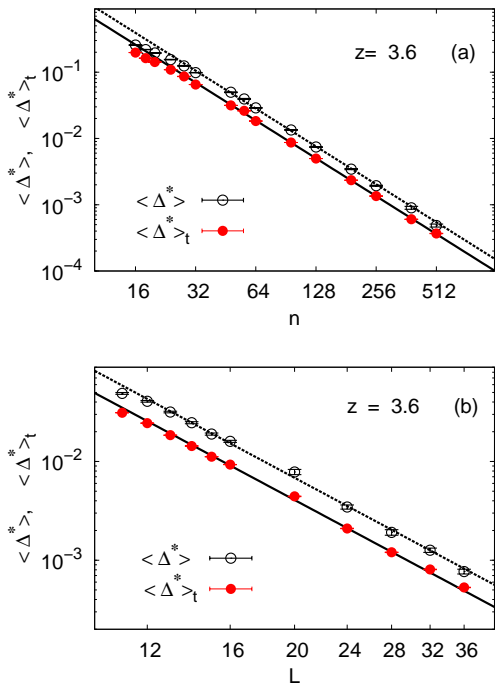}
\caption{(Color online) Finite-size scaling of the average $\langle \Delta^*\rangle $ and typical 
$\langle \Delta^*\rangle_t$ gap upper-bound for $S=0$ ($n_A=n_B$) clusters. The top and bottom panels 
show results for fixed-$n$ and $L\times L$ clusters, respectively. The lines correspond to the scaling
expected with dynamic exponent $z=3.6$ (i.e., the size dependence is $\sim n^{-z/D_f}$ and $\sim L^{-z}$, 
respectively, for the two types of clusters).}
\label{z-scale}
\end{center}
\vskip-3mm
\end{figure}

Fig.~\ref{z-scale} shows the size dependence of the disorder averaged $\langle \Delta^*\rangle$ on log-log scales 
for both fixed-$n$ (top panel) and $L\times L$ (bottom panel) clusters. We also show the typical values 
$\langle \Delta^*\rangle_t$, obtained by averaging $\ln(\Delta^*)$ for the individual clusters. While the typical 
and average values do not exactly coincide, for large systems they scale in the same way.  Linear fits to the 
$\langle \Delta^*\rangle_t$ data on the log-log scales gives $z= 3.6 \pm 0.1$ for both types of clusters. Here the 
estimated error reflects the purely statistical errors of the line fits in combination with small variations 
depending on what range of system sizes are included.

\begin{figure}
\begin{center}
\includegraphics[width=6.5cm, clip]{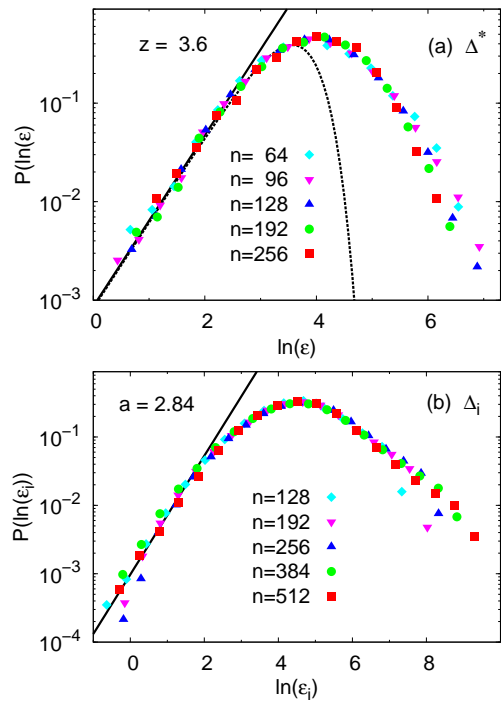}
\caption{(Color online) Distribution of the logarithm of the scaled gap upper-bound $\epsilon=\Delta^*n^{z/D_f}$ (a)
and local gap bound $\epsilon_i=\Delta_in^{a/D_f}$ (b) for fixed-$n$ clusters with ground state spin $S=0$. The exponents 
are indicated in the panels. The solid lines correspond to small-gap exponent $\omega=1$ and the curve in (a) is
a Frechet form.}
\label{gapscalingn}
\end{center}
\vskip-3mm
\end{figure}

\begin{figure}
\begin{center}
\includegraphics[width=6.5cm]{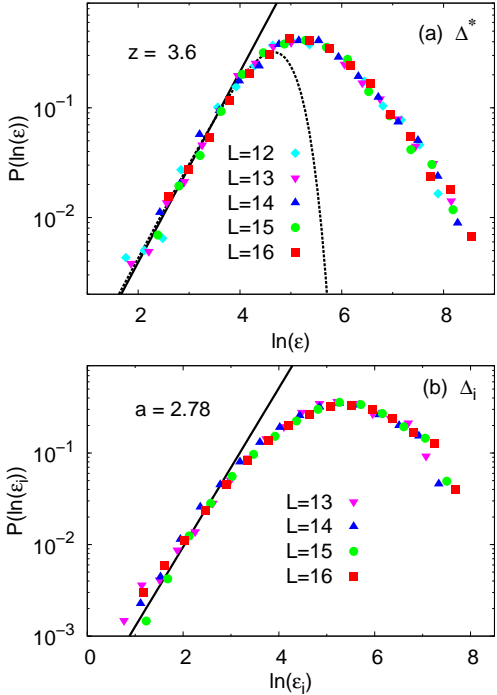}
\caption{(Color online) Probability distribution of the logarithm of the scaled gap upper-bound $\epsilon=\Delta^*L^{z}$ 
(a) and local gap bound $\epsilon_i=\Delta_iL^{a}$ (b) for $L\times L$ clusters with ground state spin $S=0$. The solid 
lines correspond to $\omega=1$ and the curve in (a) is a Frechet form.}
\label{gapscalingl}
\end{center}
\vskip-3mm
\end{figure}

As shown in Figs.~\ref{gapscalingn}(a) and \ref{gapscalingl}(a), for fixed-$n$ and $L\times L$
clusters, respectively, not only do the averages and typical values of $\Delta^*$ scale with
the system size, but the entire distribution can be collapsed onto a common size-independent curve, 
by scaling the gap estimates with the cluster size. We define the scaled gap upper-bounds for the 
two types of clusters according to
\begin{eqnarray}
\epsilon=\left\lbrace \begin{array}{ll}
\Delta^* L^z,~~~\hbox{(fixed-$n$ clusters)}, \\
\Delta^* n^{z/D_f},~~~\hbox{($L\times L$ clusters)}.
\end{array}\right.
\end{eqnarray}
As can be seen in the figures, the small-gap side of the distribution of $\ln(\epsilon)$ is very 
well described by a power law; $P[\ln(\epsilon)] \propto \epsilon^{\omega +1}$, with $\omega=1$. 
This distribution of the logarithm of $\epsilon$ corresponds to a probability distribution 
$P(\epsilon) \sim \epsilon^\omega$ for the scaled gap $\epsilon$ itself [since the differential
$d\ln(\epsilon)=d\epsilon/\epsilon$].

We next consider the local gap estimate $\Delta_i$, i.e., the inverse local susceptibility 
(\ref{localx}). Measuring this quantity at each site, we define size-scaled local gaps;
\begin{eqnarray}
\epsilon_i=\left\lbrace \begin{array}{ll}
\Delta_i L^a,~~~\hbox{(fixed-$n$ clusters)}, \label{epiloni} \\
\Delta_i n^{a/D_f},~~~\hbox{($L\times L$ clusters)}.
\end{array}\right.
\end{eqnarray}
The probability distributions of $\ln(\epsilon_i)$ for different cluster sizes, based on several 
hundred clusters of each size, collapse onto each other for a suitably chosen $a \approx 2.8$, as 
shown in Figs.~\ref{gapscalingn}(b) and \ref{gapscalingl}(b) for the two types of clusters. The
small-gap tails of the distributions are again very well described by a power law; $P(\epsilon_i) 
\sim \epsilon_i^\omega$, with the same $\omega=1$ as for the scaled ``global'' gap bound $\Delta^*$.

\subsection{\label{extremal}Extremal-value analysis}

In Ref.~\onlinecite{lingPRL2006} we used extremal value statistics~\cite{extremebook} (in a way 
generalizing a treatment of localized excitations by Lin et al.\cite{Lin}) and found a relationship 
between the exponents $z,a$, and $\omega$. For completeness, we repeat and further clarify our arguments 
here. 

Our hypothesis is that, for a large cluster of size $n$, there is a number $\propto n$ of regions of sublattice 
imbalance. These regions act as localized magnetic moments, which interact weakly with each other through the 
magnetically inert parts of the percolating cluster. The excitations of this effective low-energy system 
of coupled moments are not localized because several distant moments can be involved. It is then natural to expect 
some size dependence of the local gaps, due to the dependence of the effective interactions on the distance between the 
moments involved in a low-energy excitation, combined with the increasing distance (on average) between these moments with 
increasing cluster size. We posit that this size dependence can be captured by the single exponent $a$ in Eq.~(\ref{epiloni}).

\begin{figure}
\begin{center}
\includegraphics[width=6.5cm]{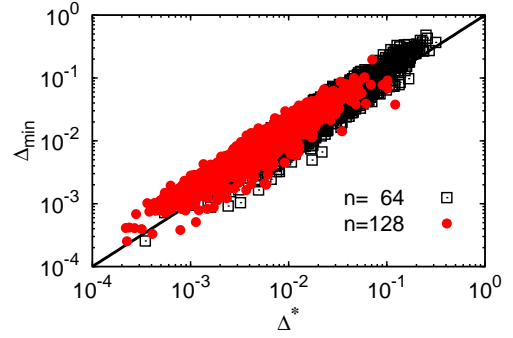}
\caption{(Color online) Correlations between the smallest local gap $\Delta_{\rm min}$ and the gap upper bound 
$\Delta^*$ for clusters of size $n=64$ and $128$. Each data point corresponds to Monte Carlo results for a randomly
generated $n_A=n_B$ cluster. The line shows the ideal (single-mode) case of complete equivalence of the two estimates 
of the finite-size gap.}
\label{gapcorr}
\end{center}
\vskip-3mm
\end{figure}

The actual finite-size gap $\Delta$  for a given cluster should correspond to the smallest of the local gaps 
$\Delta_i$ for that cluster, for which we use the notation $\Delta_{\rm min}$. Of course, the local gaps that 
we measure are only approximations; one cannot unambiguously define a local gap in an interacting system. Nevertheless, 
$\Delta_i$ reflects the local distribution of spectral weight, and there should be some site $i$ within the 
regions affected by the lowest excitation for which $\Delta_i=\Delta_{\rm min}\approx \Delta$ (and $\Delta_{\rm min} 
\ge \Delta$). In our numerical analysis, $\Delta$ is approximated by the bound $\Delta^*$, and we expect 
$\Delta_{\rm min} \approx \Delta^*$. Examining the numerical data, we indeed find a very strong correlation between 
the two quantities, as shown in Fig.~\ref{gapcorr} for fixed-$n$ clusters. Here it can be seen that $\Delta_{\rm min}$ is typically 
$1.5-2$ times larger than $\Delta^*$, which reflects larger spectral weight above the true lowest excitation 
energy in the local dynamic structure factor $S_i(\omega)$ than in $S(\pi,\pi,\omega)$. It should be noted that
$\Delta_{\rm \min} < \Delta^*$ is allowed within the sum-rule approach, although $\Delta_{\rm \min} \ge \Delta$ 
has to hold strictly. 
 
We now assume that there is a number $M \propto n$ of different local scaled gaps $\epsilon_i$ and investigate the consequences 
of this in light of the scaling behavior found above. We assume a probability distribution $P(\epsilon_i)=A\epsilon_i^\omega$ 
for some window of small $\epsilon$ (where $A$ is a constant and we consider a more general case than just $\omega=1$ 
extracted from the finite-size scaling of the data). We derive the probability distribution for the smallest scaled local gap 
$P_M(\epsilon_{\rm min})$ for large $M \sim L^{D_f}$ using extremal-value statistics. 

We should clarify why we assume $M \propto n$ for the number of local gaps, instead of just $M=n$, which is the actual number of 
different numerical values $\Delta_i$ that we compute for a given cluster. The distinction will not matter in the analysis, 
but it has an important physical significance. In our scenario, a cluster consists of regions with localized moments, which 
participate in the low-energy excitations, as well as inert parts which have only high-energy excitations. The form of the 
probability distribution $P(\epsilon_i)=A\epsilon_i^\omega$ should only hold for sites $i$ within the moment regions. It is then 
important in our analysis that also the number of such sites scales as $n$ (although one could also generalize to $M\sim n^\gamma$
with $\gamma < 1$, but the consistency of our analysis with $M \propto n$ will show that this is not necessary). Later, we will 
provide more concrete proof that the moment regions are finite and the total number of spins belonging to moments grows 
linearly with $n$.

Note also that in reality the distribution of local gaps must be cut off (equals zero exactly) below some very small value for 
a given finite cluster size. However, this should not affect the results of the analysis to follow, because also the assumed 
power-law probability is very small below such a threshold. We thus expect the results derived below to be valid within some 
significant window of scaled gaps $\epsilon$.

We denote the probability of finding a local gap at an arbitrary chosen site (within one of the moment regions) smaller than some 
value $x$ by $P_<(x)$. It is given by
\begin{equation}
 P_<(x)=\int_0^{x}P(\epsilon_i)d\epsilon_i=\frac{A}{\omega+1}x^{\omega+1}.
\label{localgap-distribution}
\end{equation}
If one of the scaled gaps $\epsilon_j$ is the smallest and has the value $\epsilon$, then all the other ($M-1$ different) values 
$\epsilon_i$, $i\not = j$ must be larger than $\epsilon$. The probability of these $M-1$ values being smaller than $\epsilon$ 
is  $[1-P_<(\epsilon)]^{M-1}$. Since any of the $M$ values could be the smallest one, we get a factor of $M$, and finally the 
distribution of the $\epsilon_j$ value is given by $P(\epsilon_j)$. Thus, the distribution of the smallest scaled local gap is
\begin{equation}
P_M(\epsilon)=MP(\epsilon)[1-P_<(\epsilon)]^{M-1},
\end{equation}
which for small $\epsilon$ also can be expressed as;
\begin{equation}
 P_M(\epsilon) = -\frac{d}{d\epsilon}{\lbrack 1-P_<(\epsilon)\rbrack}^{M}
\simeq -\frac{d}{d\epsilon}e^{-MP_<(\epsilon)}.
\label{pm-distribution}
\end{equation}
Using Eq.~(\ref{localgap-distribution}) here gives the Frechet distribution,\cite{extremebook}
\begin{equation}
P_F(u)=A{u}^{\omega}{\rm exp}[-A(\omega +1)^{-1}{u}^{\omega+1}],
\label{u-distribution}
\end{equation}
where $u=u_0\epsilon$. Thus, the probability distribution of the scaled global gap should be governed by the same exponent 
$\omega$ as the scaled local gaps. The Frechet form can indeed be fitted to the $\Delta^*$ data in Figs.~\ref{gapscalingn}(a) 
and \ref{gapscalingl}(a), with the same exponent $\omega=1$ as in the local-gap (b) panels, but only in the small-gap region. 
One cannot expect the Frechet distribution to work for large gaps, since the local gap distribution we started from is linear 
only in the small $\epsilon_i$ region (and, as discussed above, we expect the large-gap part of the distribution to be dominated
by excitations of the magnetically inert cluster regions without moments). The fitted forms in Figs.~\ref{gapscalingn}(a) 
and \ref{gapscalingl}(a) are therefore also not normalized. Nevertheless, it is encouraging that the data is in agreement with the 
result that both the local and global gaps should scale with the same exponent, which here is $\omega=1$.

\begin{figure}
\begin{center}
\includegraphics[width=6.5cm]{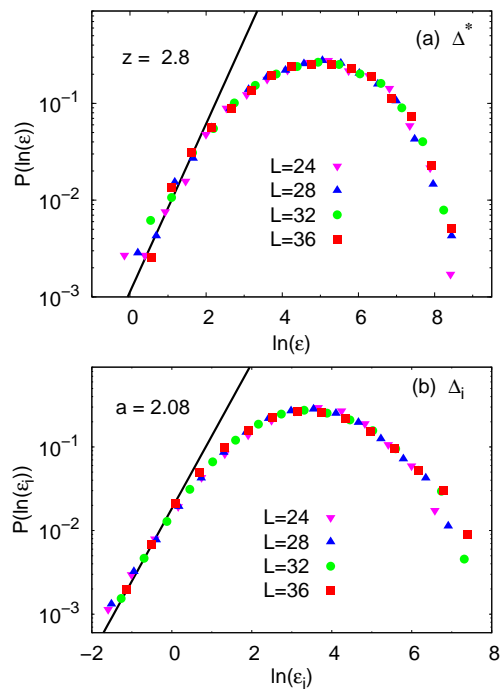}
\caption{(Color online) Distribution of the scaled gap upper-bound (a) and local gaps (b) of $L\times L$ clusters 
with no restriction on the sublattice occupation numbers $n_A$ and $n_B$. The solid line shows the asymptotic 
small-gap behavior expected with $\omega=1$.}
\label{lall}
\end{center}
\vskip-3mm
\end{figure}

Let us now use the distribution (\ref{localgap-distribution}) in a different way. Since we assume that there are 
$M \propto n$ local gaps, the typical smallest gap should correspond to $P_<(x)$ for which $x=M^{-1}$, i.e., 
$x \propto L^{-D_f}$. This gives $\Delta_{\rm min} \propto L^{-a-D_f/(\omega+1)}$.  Since $\Delta_{\rm min}$ should equal 
$\Delta$, and, by definition, $\Delta \propto L^{-z}$, we arrive at the following relationship between the three 
exponents;
\begin{equation}
z=a+\frac{D_f}{\omega+1}.
\label{zrelation}
\end{equation}
This generalizes the relation $z=D_f/(\omega + 1)$ used as a criterion for a localized excitation by Lin 
et al.~\cite{Lin} to excitations originating from two or more finite entangled regions distributed over the cluster.
With our numerical values from the finite-size scaling above, $z\approx 3.6$ and $\omega=1$ (the latter of which 
is not based on a fit, but is a value consistent with all our results), we obtain $a \approx 2.65$, in very reasonable 
agreement with the value $a \approx 2.8$ obtained in Figs.~\ref{gapscalingn}(b) and \ref{gapscalingl}(b) from the 
scaling of the $\epsilon_i$ data for the fixed-$n$ and $L\times L$ clusters. The applicability of the exponent relation 
(\ref{zrelation}) provides strong support to our hypothesis of ``globally entangled local moment excitations''.

\subsection{\label{generals} SSE results for general-$S$ clusters}

\begin{figure}
\begin{center}
\includegraphics[width=6.5cm]{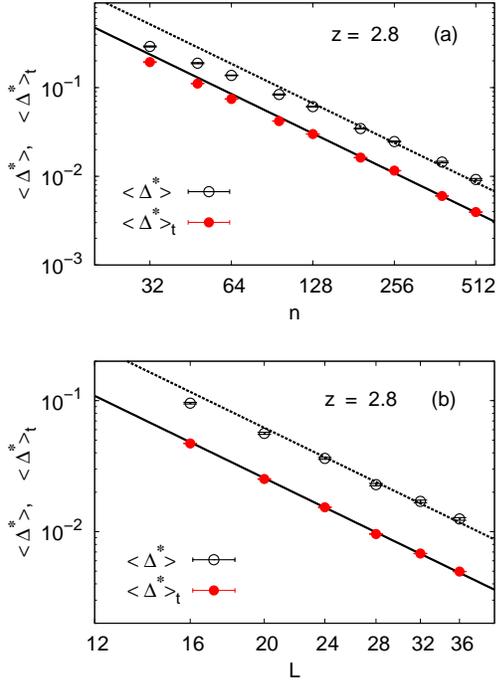}
\caption{(Color online) Finite-size scaling of the average and typical gap upper-bound $\Delta^*$ for the ensemble
with unrestricted $n_A$ and $n_B$. The upper and lower panels show results for fixed-$n$ and $L\times L$ clusters, 
respectively. Lines corresponding to a dynamic exponent $z=2.8$ are shown with all the data sets.}
\label{z-scale-2}
\end{center}
\vskip-3mm
\end{figure}

We now turn to clusters with no restriction on the sublattice occupations $n_A$ and $n_B$ in the generated ensemble.
Scaling results for the global and local gaps of $L\times L$ clusters are shown in Fig.~\ref{lall}. The finite-size 
scaling of the average and typical values of $\Delta^*$ are shown for both fixed-$n$ and $L\times L$ clusters in 
Fig.~\ref{z-scale-2}. We obtain $a\approx 2.1$ and $z\approx 2.8$ for both cluster types. These exponents differ 
significantly from the ones obtained previously for the ensemble including $S=0$ clusters only. In particular, 
$z\approx 1.5D_f$, whereas the $S=0$ clusters gave $z \approx 2D_f$. The exponent relationship 
(\ref{zrelation}) still holds approximately, albeit with somewhat larger deviations than in the $S=0$ case. 
The small-gap behavior remains consistent with the exponent $\omega=1$ in all cases. 

We believe that the much smaller exponents $z,a$ are due to a failure of the sum rule approach to capture the true 
low-energy states for $S>0$. To demonstrate this, we next investigate the dynamic structure factor (\ref{dynamicsS}). 

\section{\label{spectralsection}Spectral weight distribution}

We will investigate how the spectral weight of the dynamic structure factor is distributed among different spin sectors 
of the excited states in (\ref{dynamicsS}). Acting on the ground state with the ${\bf q}=(\pi,\pi)$ spin operator (\ref{sqop}), or 
the corresponding $x$ or $y$ components, on one of the $(2S+1)$ degenerate ground states of spin $S$ results in states 
with spin $S$ and $S\pm 1$. This well known selection rule for the dynamic structure factor (\ref{dynamicsS}) can be 
easily demonstrated in the valence bond basis. Here we do this as a prelude to discussing the distribution of the spectral 
weight among the three sectors of final spin for clusters with ground state $S>0$.

\subsection{Selection rules}

\begin{figure}
\begin{center}
\includegraphics[width=6.25cm]{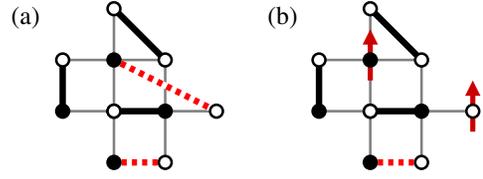}
\caption{(Color online) Valence bond states on a cluster with sublattice imbalance $\Delta_{AB}=1$ (requiring one
non-bipartite bond---here the top one). Open and solid circles indicate the two sublattices. Singlet and triplet bonds are 
shown as solid and dashed lines, respectively. (a) shows a state with two triplet bonds and $m_z=0$. In (b), there are
two unpaired up spins and $m_z=1$.}
\label{vbconfigs}
\end{center}
\vskip-3mm
\end{figure}

We consider an extended valence bond basis with an arbitrary number of $m_z=0$ triplet bonds (\ref{psitriplet}) in addition to 
singlet bonds, for a state with total $m_z=0$ (hence the number of spins, $N$, is even). Later, we will consider also $m_z\not =0$. 
The standard valence bond basis for $n_A=n_B$ is restricted to bipartite bonds only.\cite{liang2} Here, for $n_A\not=n_B$ and $m_z=0$, 
we require a maximal number of bipartite bonds, i.e., if the total sublattice imbalance is defined as $\Delta_{AB}=|n_A-n_B|/2$, there 
will be $n_b=N/2-\Delta_{AB}$ bipartite bonds and $n_c=\Delta_{AB}$ bonds connecting sites on the same sublattice (with all such pairs 
either on the A or B sublattice, depending on which sublattice has the larger number of sites). This basis is clearly overcomplete. 
A state with two triplet bonds and one non-bipartite bond is illustrated in Fig.~\ref{vbconfigs}(a). 

First, let us discuss the relationship between the total spin $S$ and the number of triplet bonds. A state with $n_t$ 
triplets does not have fixed spin when $n_t>1$ (while for $n_t=0$ and $1$, the state has fixed $S=0$ and $1$, 
respectively). According to the rules for addition of angular momenta, one might at first sight suspect that $n_t$ 
triplets could be used to form states with $S=0,1,\ldots,n_t$. However, consider the operator $Z$ which inverts 
all the spins;
\begin{equation}
Z|S^z_1,S^z_2,\ldots,S^z_N\rangle=|-S^z_1,-S^z_2,\ldots,-S^z_N\rangle,
\end{equation}
which is a special case of a rotation in spin space. Since the total magnetization $m_z=0$, a state with fixed $S$ is 
also an eigenstate of this operator, with eigenvalue $z=\pm 1$. Since a triplet pair (bond) is even under $Z$ while a singlet 
pair is odd, the eigenvalue $z$ of a state with a fixed number $n_t$ of triplet bonds is $z=(-1)^{N/2-n_t}$. Thus, in order to 
construct a state with fixed $S$ (fixed $z$), one cannot mix valence bond states with even and odd number of triplets. Since the minimum 
number of triplets required to construct a state with fixed spin is $n_t=S$, we conclude that the triplet numbers 
that can be mixed are $n_t \in \{S,S+2,\ldots N/2\}$. This, in turn, implies that a state with fixed number of triplets is 
a linear combination of states with $S \in \{0/1,\ldots,n_t-2,n_t\}$, where the lower limit $0$ or $1$ applies for even 
and odd $S$, respectively.

Next, we let the $q=(\pi,\pi)$ spin operator (\ref{sqop}) act on a given valence bond state with $n_t$ triplets. 
We can write the operator in a way tailored specifically for the state under consideration;
\begin{equation}
S^z_{\pi,\pi} = \frac{1}{\sqrt{N}} \left [\sum_{b=1}^{n_b} (S^z_{i(b)}-S^z_{j(b)}) +
\sum_{c=1}^{n_c} (S^z_{k(c)}+S^z_{l(c)}) \right ].
\label{szbipartite}
\end{equation}
Here the subscripts $i(b)$ and $j(b)$ refer to two sites connected by a bipartite valence bond $b$, and $k(c),l(c)$ denotes 
a pair of sites on the same sublattice, connected by a non-pipartite bond $c$. The bonds can be singlets or triplets, and the 
possible outcomes when operating with one of the terms are;
\begin{eqnarray}
& & (S^z_{i(b)}-S^z_{j(b)})|...(i_b,j_b)... \rangle =|...[i_b,j_b]... \rangle, \nonumber \\
& & (S^z_{i(b)}-S^z_{j(b)})|...[i_b,j_b]... \rangle =|...(i_b,j_b)... \rangle, \nonumber \\
& & (S^z_{i(c)}+S^z_{j(c)})|...(i_c,j_c)... \rangle =0,\\
& & (S^z_{i(c)}+S^z_{j(c)})|...[i_c,j_c]... \rangle =0.\nonumber
\end{eqnarray}
Thus, operating with the full $S^z_{\pi,\pi}$, we obtain a linear combination of states with $n_t+1$ and $n_t-1$ triplets.
Extending this result to the case of a  fixed-$S$ state $|\Psi_S\rangle$, which is a linear combination of states 
with different $n_t$ (all even or all odd), we can think of the triplet bond created or destroyed in each term [with
the operator (\ref{szbipartite}) written in the appropriate way for operation on each term] as adding or subtracting
a spin $1$ to or from a spin $S$. Then, considering also that even and odd $S$ corresponds to mixtures of even and odd 
$n_t$, respectively, we conclude that the state $S^z_{\pi,\pi}|\Psi_S\rangle$ is a mixture of only $S\pm 1$ states (which 
is also consistent with the fact that for $S=0$ ground states, the spectral weight is exclusively due to $S=1$ excitations).

In order to respect the spin-rotational invariance when using the $z$-component operator $S^z_{\pi,\pi}$ in the dynamic
structure factor for $S>0$, we also have to consider non-zero $m_z$. Some of the spins are then not paired up into valence bonds. 
In a minimal basis mixing valence bonds and spins, there are $2m_z$ unpaired up or down spins for $m_z>0$ and $m_z <0$, respectively. 
The unpaired spins cannot be restricted to the same sublattice, so now the basis consists of the unpaired spins at arbitrary locations, 
a maximal number of bipartite bonds on the remaining locations, and the rest of the sites covered by non-bipartite bonds. 
An example of such a state is illustrated in Fig.~\ref{vbconfigs}(a). 

In Eq.~(\ref{szbipartite}) $n_b$ and $n_c$ are the number of bipartite and non-bipartite bonds in a given basis state and $n_b+n_v=n-m_z$.
We now also have to add a sum over the $2m_z$ unpaired spins. It is then clear that $S^z_{\pi,\pi}|\Psi_S\rangle$ will contain also 
a spin-$S$ component, arising from this added sum, in addition to the $S\pm 1$ components (which can be argued for in analogy 
with the $m_z=0$ case). Some, but not all, of the corresponding spectral weight in the spin $S$ sector is at $\omega=0$, as discussed 
in Sec.~\ref{ssemethod}. Averaging over all $m_z = -S,\ldots,S-1,S$, it is also clear that the amount of $S \to S$ spectral 
weight should increase with $S$, as it is zero for $S=0$ and the relative weight of the operations on unpaired spins 
increases with $m_z$.

\subsection{Results for small clusters}

\begin{figure}
\begin{center}
\includegraphics[width=6.5cm]{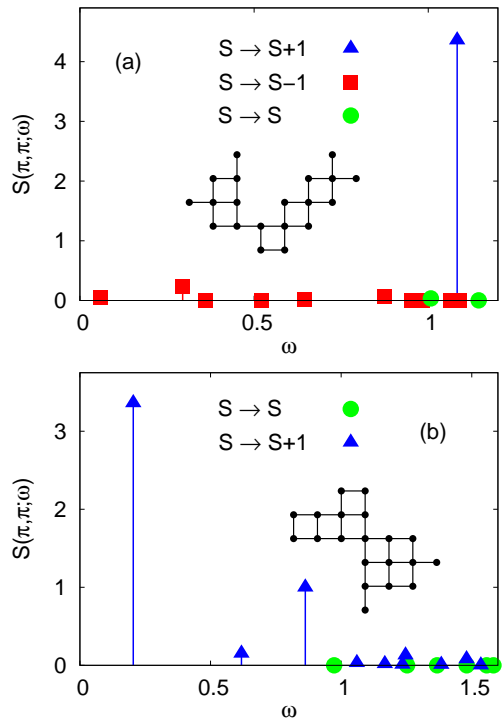}
\caption{(Color online) Dynamic structure factors of two $20$-site clusters
  with ground state spin $S=3$ (a) and $S=0$ (b). The cluster shapes are drawn in
  the panels. The delta-functions in Eq.~(\ref{dynamicsS}) are represented by vertical
  lines of length equaling the spectral weight. The symbols on top of the line indicate
  the spin of the corresponding excited states relative to the ground state $S$. In (b), there
  is no $S \to S$ spectral weight; the circles only indicate the locations of such states.} 
\label{spectrum}
\end{center}
\vskip-3mm
\end{figure}

We now turn to numerical results for the dynamic structure factor. Investigating small clusters with the Lanczos method, 
we have found that the lowest excitation of a cluster with ground state spin $S$ almost always has spin $S-1$, whereas the dominant 
spectral weight arises from a state with $S+1$. An example of this behavior is shown in Fig.~\ref{spectrum}(a) for a cluster 
with ground state spin $S=3$. The spectrum is dominated by a large contribution from an $S=4$ state at $\omega/J \approx 1$. 
However, there are numerous very small contributions from $S=2$ states below this peak, including the lowest excitation 
at $\omega/J\approx 0.05$. In this case the sum rules give a bound $\Delta^*$ very close to the energy of the 
lowest $S=4$ state, and, thus, differs from the true gap $\Delta$ by a factor of $20$. In contrast, Fig.~\ref{spectrum}(b)
shows results for a cluster with $S=0$ ground state. Here there are of course no excitations with $S-1$, and all the 
spectral weight is in the $S+1=1$ channel. Moreover, the dominant weight originates from the lowest excitation. The sum 
rule approach here gives a bound reasonably close to the true gap.

\begin{figure}
\begin{center}
\includegraphics[width=6cm]{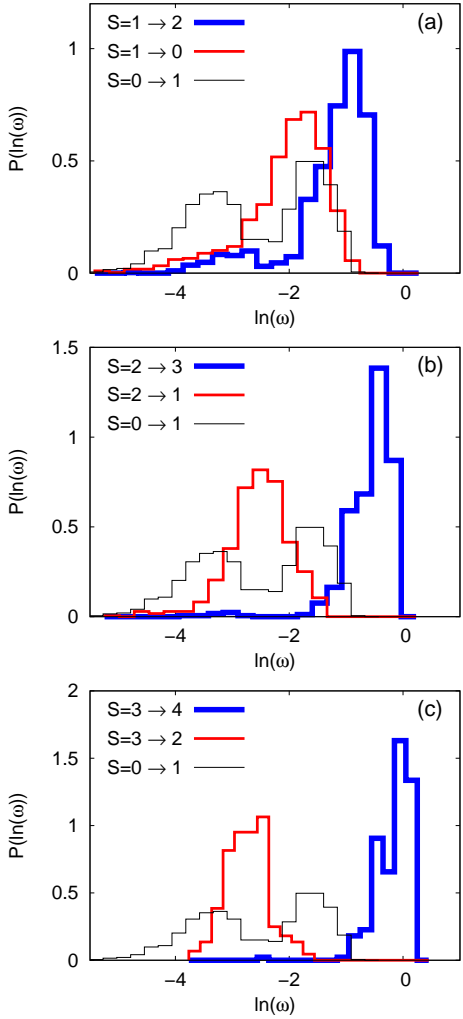}
\caption{(Color online) Distribution of the energies of the lowest excitation with $S=S\pm 1$ in $n=20$ clusters with ground 
state spin $S$. Results for $S=1,2,3$ are shown in panels (a),(b),(c). In all panels, results for the lowest triplet excitation 
of $S=0$ clusters is shown for comparison.}
\label{lanc-histo}
\end{center}
\vskip-3mm
\end{figure}

We further examine the statistics of the gaps corresponding to excitations with $S\pm 1$ for $n=20$ clusters with ground state $S$. 
In Fig.~\ref{lanc-histo} we show histograms based on several hundred clusters with $S=1,2,3$, along with results for $S=0$ clusters 
for comparison. We can see that the distribution of the $S+1$ excitations is peaked at higher energies than the $S-1$ ones, and the 
distance between the two distributions grows with $S$. As we discussed in Sec.~\ref{moredata_a}, the distribution of singlet-triplet 
excitation gaps is double-peaked for small systems, with the upper peak diminishing as a function of the cluster size. In 
Fig.~\ref{lanc-histo} the lower part of the $S=0 \to 1$ distribution is located below the $S-1$ distributions for $S>0$ clusters.
It appears plausible from these results that the $S \to S-1$ and $S \to S+1$ gaps can have different scaling properties.

It is also clear from these calculations that the sum rule approach for $S>0$ clusters does not reflect the true smallest gaps, 
which are due to $S-1$ excitations, but instead reflect the 
distribution of spectral weight of $S+1$ excitations. The quantity $\Delta^*$ therefore has
a different meaning, which can still be physically relevant because many experimental techniques 
probe $S(q,\omega)$ directly, e.g., neutron scattering and nuclear magnetic resonance. These
experiments should observe low-energy dynamics corresponding to $z\approx 1.5D_f$, according
to our results in the previous section. The most plausible scenario is that the lowest $S-1$ excitation energies, 
for large clusters and typical ground state spin (which is of the order $\sqrt{n}$), scale with the same 
dynamic exponent $z\approx 2D_f$ as the triplet excitations of $S=0$ clusters (which we will argue further
also in the next section). While their low spectral weights would make them difficult to observe in measurements
sensitive to $S(q,\omega)$, they are of course still relevant for thermodynamic properties such as the specific heat.

\begin{figure}
\begin{center}
\includegraphics[width=6cm]{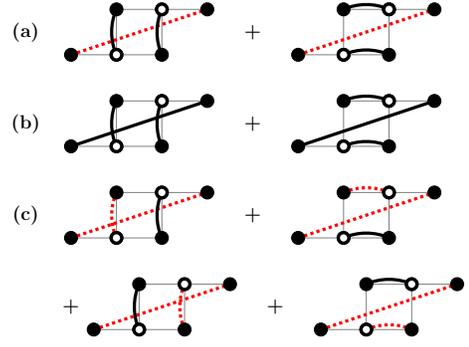}
\caption{(Color online) Valence bond states corresponding closely to true eigenstates of a
6-site cluster with ground state spin $S=1$. Solid and dashed bonds correspond to singlets and triplets,
respectively. (a) is the ground state, (b) the lowest $S=0$ excitation, and (c) is obtained from (a) by
acting on it with $S^z_{\pi,\pi}$ (which is a good approximation to the lowest $S=2$ excitation).}
\label{cluster6}
\end{center}
\vskip-3mm
\end{figure}

\subsection{\label{vbtheory}Valence bond theory}

We now address the important issue of why the $S \to S-1$ contribution to the spectral weight is so small. We will
argue that this is, in fact, consistent with our scenario of the low-energy excitations being due to effectively isolated 
magnetic moments. To illustrate this point, Fig.~\ref{cluster6} shows valence bond states for a 6-site cluster with 
ground state spin $S=1$. The state in (a) is constructed as an approximate ground state based on the notion that triplet 
bonds should be predominantly located in regions of sublattice imbalance. This cluster has two ``dangling'' spins, 
which we take at maximum separation. For the two singlet bonds, we construct a symmetric combination (which corresponds 
to the true ground state of the Heisenberg model on the four sites in isolation). It is now natural to assume that the lowest 
excitation corresponds to converting the triplet bond into a singlet, as shown in (b). Diagonalizing the hamiltonian 
exactly, we find that these simple states indeed are good approximations to the eigenstates; the overlap of (a) with 
the true ground state is $0.814$, while the overlap of (b) with the lowest $S=0$ state (which is the lowest excited 
state) is even larger, at $0.973$. On the other hand, if we act with $S^z_{\pi,\pi}$ on state (a), as explained above 
with the spin operator written in the form (\ref{szbipartite}), we obtain the state shown in Fig.~\ref{cluster6}(c).
This state mixes $S=0$ and $S=2$ states, and its overlap with the actual lowest $S=2$ state is $0.789$. The overlap 
of (c) with the approximate $S=0$ state (b) is exactly $0$, and the overlap with the exact lowest singlet also 
vanishes. In the case of the (b),(c) overlap, it is immediately clear that it is zero because of their different
states of the long bond. The states also differ in the quantum number related to a $180^\circ$ rotation of the
cluster; (b) is odd and (c) even under this symmetry transformation. The true ground state is also odd, which
explains why the overlap with state (c) is exactly $0$. This latter property is of course particular to this symmetric 
6-site cluster. In general, for a less symmetric larger cluster with two dangling spins, we would expect some small overlap 
between $S^z_{\pi,\pi}|\Psi_S\rangle$ and the lowest $S-1$ state, because the triplet will not be exactly localized 
at only two sites

Based on the above example, we can understand that, in general, $S>0$ ground states contain some triplet bonds
connecting non-bipartite sites. The lowest excitation should normally have spin $S-1$ and closely correspond to 
converting one triplet bond into a singlet. On the other hand, acting with the spin operator one obtains a linear 
combination of $S-1$ and $S+1$ states with an additional triplet bond, and the overlap of the $S-1$ component with the 
low-energy states with this spin is low (because of the differing singlet/triplet state of one non-bipartite bond). The lowest 
$S-1$ excitation should thus be very similar to the excitations we have argued for in the case of the $S=1$ excitations 
of a singlet ground state, which essentially corresponds to promoting a long singlet (between two moments, which can be 
located far away from each other) into a triplet. For an $S>0$ cluster we instead demote a long triplet bond into a 
singlet. This similarity also suggests that the true dynamic exponent (giving the scaling of the lowest energy, not the 
dominant spectral weight) in the case of $S>0$ clusters should be the same $z\approx 2D_f$ that we have found for the $S=0$ 
clusters.

\section{\label{bilayerp}Bilayer model at $p^*$}

Our hypothesis for the low-energy excitations is that they are due to effectively unpairable 
spins on the percolating cluster. To test this hypothesis further, we consider a case where
there are no such spins; the bilayer Heisenberg antiferromagnet with ``dimer dilution'', i.e., 
two identical clusters coupled through a nearest-neighbor inter-layer coupling $J_\perp=gJ$.
The hamiltonian for this system was already written down in Eq.~(\ref{hambilayer}). Its static
properties were studied in Refs.~\onlinecite{andersPRL2002,vajkPRL2002,vojtaPRB2006,andersPRL2006}. 
The percolating cluster remains ordered at $T=0$ when the coupling ratio $g \alt 0.1$, whereas for 
larger inter-layer couplings the cluster is quantum disordered. Here we consider $g=0.01$; well inside 
the ordered regime. One might then expect the quantum rotor picture to be valid, as has been argued also 
based on field theoretical considerations,\cite{vojtaPRL2005} and, thus, the dynamic exponent should 
be $z=D_f\approx 1.89$. Scenarios, involving ``fractons'' are also possible.\cite{orbachModernPhy1994}

\begin{figure}
\begin{center}
\includegraphics[width=7cm]{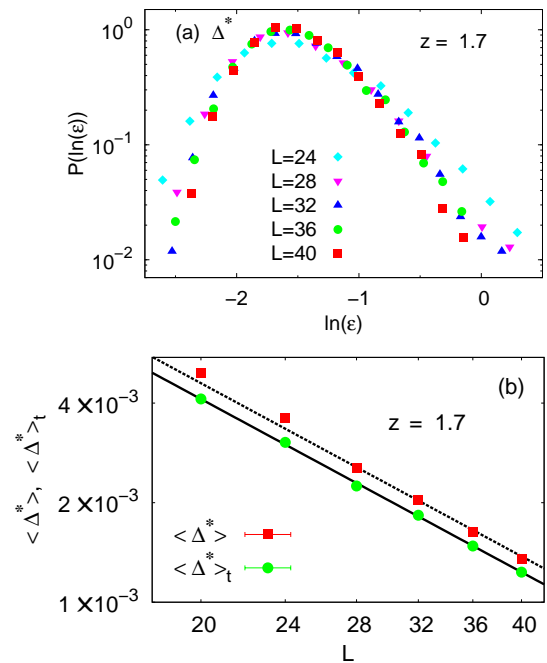}
\caption{(Color onine) Properties of the gap upper-bound $\Delta^*$ for dimer-diluted bilayer systems at 
 inter-layer coupling $g=0.01$ at the percolation point. (a) shows the scaling of the full probability 
 distribution with $z=1.7$, while (b) shows the size dependence of the average and typical values, along
 with lines corresponding to the asymptotic behavior with $z=1.7$.}
\label{bilayerscaling}
\end{center}
\vskip-3mm
\end{figure}

Fig.~\ref{bilayerscaling} shows scaling results of the kind we previously discussed
for the single layer. The peak of the probability distribution of the bound $\Delta^*$
for different cluster sizes $L$ (the largest cluster of diluted $L\times L$ lattices) 
coincides when scaled with $L^z$ and $z\approx 1.7$. This exponent is slightly 
smaller than $D_f$, but considering statistical uncertainties of several percent and effects of 
subleading size corrections, $z=D_f$ is plausible, in contrast to $z\approx 2D_f$ in the single 
layer. Note that the data in Fig.~\ref{bilayerscaling}(a) do not collapse onto a single curve 
as clearly as in the single-layer plots \ref{gapscalingn} and \ref{gapscalingl}. The scaled
gap distributions instead appear to become narrower with increasing $L$. This may be due to
self-averaging following from the global nature of quantum rotor excitations. The local gap distribution 
(not shown here) also does not scale well with $L$.

\section{\label{belowpc}Single layer away from the percolation point}

An interesting question is whether the small energy scale of the single-layer clusters at $p^*$ survives also away 
from the percolation point. We here examine $L\times L$ systems diluted at $p<p^*$, again studying the largest cluster 
for each dilution realization (which now is two-dimensional; $\langle n\rangle \sim L^2$). We only consider clusters 
with ground state spin $S=0$.

\begin{figure}
\begin{center}
\includegraphics[width=6.5cm]{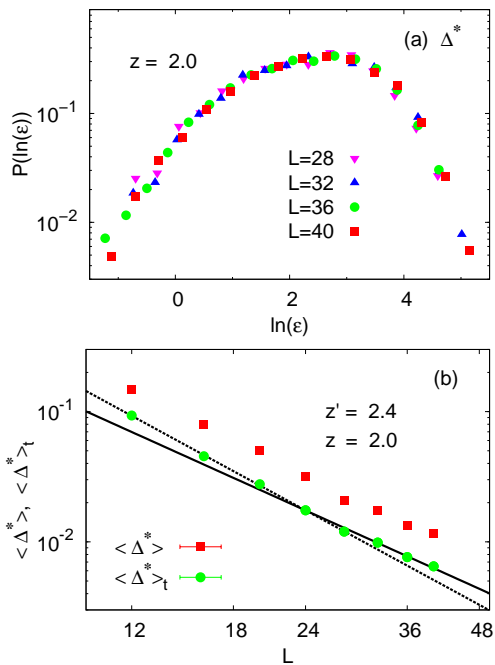}
\caption{(Color online) Scaling properties of $\Delta^*$ of single-layer clusters at dilution fraction $p=0.3$. 
For $L\ge 28$, the distribution can be collapsed with dynamic exponent $z=1.8$ as shown in (a). For smaller sizes, 
there is a cross-over behavior, as shown in (b) for the average and typical $\Delta^*$. The two lines correspond to
$z=2$ (a possible asymptotic value) and $z'=2.41$ (a pseudo-scaling exponent in a cross-over regime).}
\label{singlelayerp03}
\end{center}
\vskip-3mm
\end{figure}

Fig.~\ref{singlelayerp03} shows results for the gap upper-bound at $p=0.3$. For the largest few sizes the data 
are consistent with power-law scaling corresponding to $z=D=2$ (with a statistical error of $\approx 10\%$); very
different from the behavior at $p^*$. Given our scenario for the excitations exactly at $p^*$, the much smaller $z$ 
away from $p^*$ is either an indication of the moment regions not existing, or their mutual effective couplings 
(or their couplings to the rest of the cluster) being much stronger, thereby invalidating the picture of an effective
low-energy subsystem. We still expect regions of sublattice imbalance away from $p^*$, as we will discuss further
in the next section. It may not be surprising, however, that the moments associated with these are not weakly coupled, 
because for any $p<p^*$ the largest cluster has a finite spin stiffness (also in the thermodynamic limit), whereas 
exactly at $p^*$ the stiffness vanishes (although the cluster is still ordered).\cite{andersPRB2002} The order is thus 
much more robust, and as a consequence all the effective moments at $p<p^*$ may be locked to the global N\'eel vector 
and cannot be regarded as weakly coupled semi-independent degrees of freedom. We will discuss this further 
in Sec.~\ref{perco-summary}.

For systems very close to the percolation point, $p^*\approx 0.407$, one cannot expect to detect differences from the 
behavior exactly at $p^*$. For the smaller cluster sizes at $p=0.3$ we can observe in Fig.~\ref{singlelayerp03} what 
is likely a cross-over behavior from the behavior at $p^*$ to the asymptotic scaling behavior at $p=0.3$. The effective 
exponent below sizes $L \approx 20$ is smaller than the value we found at $p^*$, but, on the other hand, $p=0.3$ is 
already quite far away from $p^*$ and it is not surprising that a different behavior obtains here. Closer to $p^*$
we expect data for small sizes to scale with $z\approx 2D_f$, but to observe clearly this scaling, followed by a 
cross-over to $z=D=2$, would require larger clusters than we can access currently.

It should be noted that the results discussed here (and those for the bilayer in the previous section) do neither prove 
that the mapping to quantum rotors holds for $p<p^*$ (and in the bilayer at $p^*$), nor that the dynamic exponent exactly 
equals $D=2$ (or $D_f$). For fracton excitations, one would expect $z\not=D_f$ (but close to $D_f$).\cite{terao94} It would 
therefore be useful to determine $z$ for the single layer at $p<p^*$ and the bilayer at $p^*$ to higher precision, which, 
however, is a very demanding task that we leave for future studies.

\section{\label{classicalmodel}Classical dimer-monomer model}

In the mapping of a quantum antiferromagnet onto a quantum rotor model,\cite{sachdevbook} one 
assumes that there is local antiferromagnetic order on some length scale $\Lambda$. A subsystem $i$ 
of the system, of length $\Lambda$, is then replaced by a quantum rotor ${\bf L}_i$, which can 
reproduce the ``Anderson tower'' of low-energy states of different total spin $S$ (which the 
subsystem would exhibit in isolation). The rotors for all the subsystems are then coupled 
in a way consistent with the expected dominant fluctuations and symmetries of the system. 
For such a mapping to produce the correct physics, the subsystems should consist of an even number of spins, 
arranged in such a way that their ground state, in isolation, is a singlet. If the system geometry does not 
allow for such a decomposition, the situation will be more complicated. The question is then; how can one
decompose the system into quantum rotors and ``left-over'' spins in a well defined way, which maintains
the salient features of the disordered clusters?

The smallest unit for which a local quantum rotor can be considered in a hypothetical mapping is a dimer 
consisting of two nearest-neighbor spins, which in isolation has a singlet $(l=0)$ ground state and 
a triplet ($l=1$) excited state. This corresponds to a quite severely truncated rotor tower, but the 
local cut-off should not matter for the low-energy physics of the coupled system. We have already 
discussed the fact that a disordered cluster cannot normally be fully decomposed into such dimers, as 
there would in most cases be some ``dangling spins'' (or, more generally, regions of imbalance in the 
sublattice occupation numbers) left over after the cluster has been maximally covered with close-packed dimers. 
If we consider larger subsystems, there will be similar problems, i.e., not all subsystems will have 
singlet ground states in isolation. We will here proceed to investigate the geometric decomposition
of the system into nearest-neighbor dimers and left-over monomers.

In a standard classical dimer model,\cite{fisher61} a dimer corresponds to two connected nearest-neighbor sites, here on 
a square lattice. The statistical mechanics problems corresponds to counting all the dimer coverings. In a dimer-monomer 
model,\cite{fisher63} there are also some unpaired sites present, and the counting now includes all possible dimer 
and monomer configurations; normally at a fixed density of monomers. In the case at hand here, we investigate disordered 
clusters, and we want to maximally cover the clusters with dimers. There will then typically be some left over monomers 
that cannot be paired. For a given cluster, we want to sample dimer-monomer configurations with the smallest possible 
number of monomers. We are interested in the spatial distribution of monomers, which provides us with a concrete 
quantitative measure of ``sublattice imbalance''. We want to identify the regions of sublattice imbalance and 
investigate the size distribution of these regions.

\begin{figure}
\begin{center}
\includegraphics[width=5.25cm]{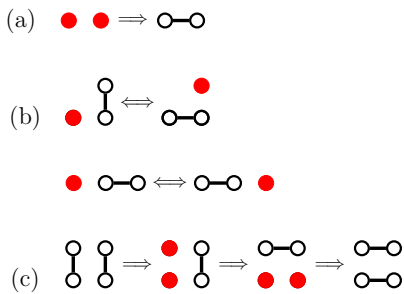}
\caption{(Color online) Updating processes used in Monte Carlo sampling of the classical dimer-monomer
aggregation model. (a) is the annihilation of two monomers, leading to a dimer. (b) shows the two
elementary monomer-dimer moves. In (c), a monomer pair is temporarily created out of a dimer. One
of the monomers is then moved until it can be annihilated with another monomer. A large number of dimers 
can be changed in such ``loop updates''.}
\label{dimermcmoves}
\end{center}
\vskip-3mm
\end{figure}

Here we consider clusters constructed on $L\times L$ lattices, with, as before, only the largest cluster found in each 
realization included in the statistics. In Monte Carlo sampling of the dimer-monomer configurations on these clusters, 
we start with an arbitrary configuration, e.g., one containing only monomers. 
The updating scheme is illustrated in Fig.~\ref{dimermcmoves}. In (a), when two monomers are 
located next to each other, they are annihilated and form a dimer. Dimers and monomers can be updated 
together according to the two moves shown in (b). We can also break a dimer into two monomers, 
as in (c). One of the monomers is then moved, together with dimers as in (b), until it encounters
a monomer (which can, but does not have to be, the same as the one it was originally paired with), together
with which it again can be combined to form a dimer as in (a). This is an efficient way to update parts of
the cluster where there are no monomers (other than the two introduced for the purpose of the
update). This simulation process will eventually converge to a state with the minimum number of monomers 
for a given cluster, because the monomer annihilation process (a) is always carried out when possible. 
Whenever two monomers are created, they will eventually be annihilated. Our model is therefore also an 
aggregation model for dimers.

For a given cluster generated on an $L\times L$ lattice, after a long equilibration to make sure that the 
minimum monomer number has been reached, we collect statistics. One quantity of interest is the average monomer 
density for each site. Most of the sites never have any monomers. We define a ``moment'' as a region consisting
of $S_m$ sites to which one or several monomers are confined. By definition, a monomer inside such a moment cannot move to a 
different moment through the Monte Carlo processes. In addition, the moments consist only of sites on the same sublattice, 
because an individual monomer only moves on a given sublattice, as in Fig.~\ref{dimermcmoves}(b). Two moments on 
different sublattices cannot have any sites that are nearest neighbors, because then two monomers in these 
different regions could become adjacent and annihilate each other. 

\begin{figure}
\begin{center}
\includegraphics[width=7.25cm, clip]{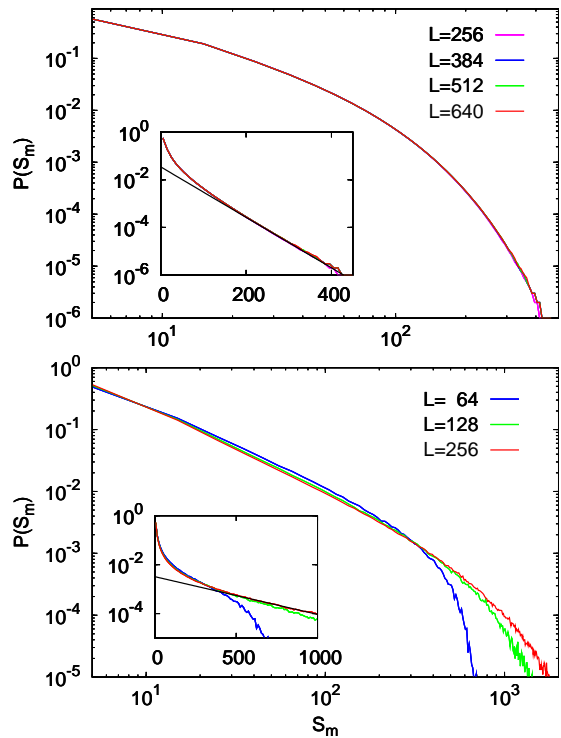}
\caption{(Color online) Probability distribution of the moment size of the classical  dimer-monomer model at the 
percolation point $p^*$ (top panel) and at $p=0.3$ (bottom panel) graphed on a log-log scale. Hundreds of realizations of the 
largest cluster on $L\times L$ lattices were used. For large $L$ the distributions collapse onto a single curve, reflecting a 
finite typical moment size. The average moment size is $\langle S_m\rangle=16$ at $p^*$ and $47$ at $p=0.3$. The asymptotic 
form of the probability distribution for large clusters is $\propto {\rm e}^{-S_m/\sigma}$, as shown in the insets using a
semi-logarithmic scale.}
\label{momentsize}
\end{center}
\vskip-3mm
\end{figure}

Keeping track of the moment regions and their sizes involves straight-forward book-keeping, and we just proceed to discuss
results. Fig.~\ref{momentsize} shows the size distribution of the moments both at the percolation point and away from it, 
at $p=0.3$. For small moment sizes $S_m$, the distribution is close to a power-law, especially 
at $p=0.3$, but there is a cross-over to a clearly exponential decay for large $S_m$. The distribution can be fitted well with 
the form ${\rm e}^{-S_m/\sigma}$, with $\sigma\approx 42$ and $\approx 300$ at $p^*$ and $p=0.3$, respectively. The average moment 
size $\langle S_m\rangle$ computed as a sum over all the sizes is smaller; $\langle S_m\rangle\approx 16$ at $p^*$ and $\approx 47$ 
at $p=0.3$. In the figure, the largest cluster sizes are much larger than $\langle S_m\rangle$ and $\sigma$, and the curves for the 
largest $L$ overlap almost completely.

These calculations prove that the notion of local sublattice imbalance is well defined and quantifiable. Finite moment regions 
exist both at and away from the percolation point. In the next section, we will present result from valence bond quantum Monte Carlo 
simulations in the triplet sector. We will there also look at the spatial distribution of the monomers in the classical dimer-monomer 
model, and compare it with the distribution of the triplet in the lowest excitation of the actual quantum spin model.

\section{\label{rvbqmc}Spatial distribution of singlet-triplet excitations}

As discussed in sec.~\ref{ipr}, the valence-bond projector QMC method offers us the possibility to
examine the lowest triplet state in a unique way. In a disordered system, the spatial distribution
of the triplet bond gives a very direct measure of the extent to which different parts of the system are 
affected when exciting a cluster with $S=0$ ground state to its lowest $S=1$ state. An example of the triplet 
density for a very small cluster was already presented in Fig.~\ref{clusterdemo}. It should be noted that 
the statistics of the singlet bonds is also affected by the presence of a triplet bond, and, thus,
just examining the properties of the triplet bond does not give a complete picture of the excitation.
However, if a large region of the system has no (or very low) average triplet density, then the singlets
of that region should also not be much affected. The spatial distribution of the triplet should therefore
provide a valid measure of the tendency (if any) of the triplet excitation to localize.

\begin{figure}
\begin{center}
\includegraphics[width=7cm, clip]{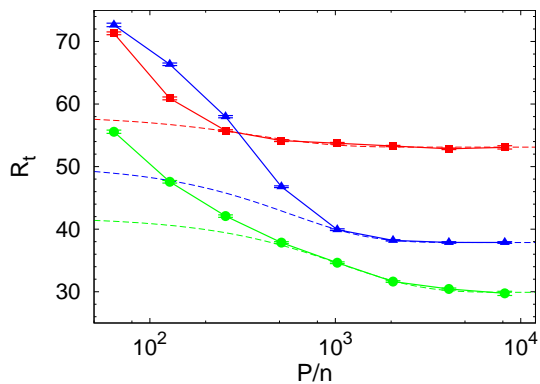}
\caption{(Color online) (a) Examples of the convergence of the average triplet IPR for three individual clusters as a function of 
the  projection power, here normalized by the cluster size as $P/n$. The clusters were generated on $26\times 26$ lattices. 
The dashed curves are of the form $a+b{\rm e}^{-cP/n}$, with $a,b,c$ adjusted to fit the last few (large $P/n$) points.}
\label{vbsimulation}
\end{center}
\vskip-3mm
\end{figure}

\begin{figure}
\begin{center}
\includegraphics[width=5.75cm, clip]{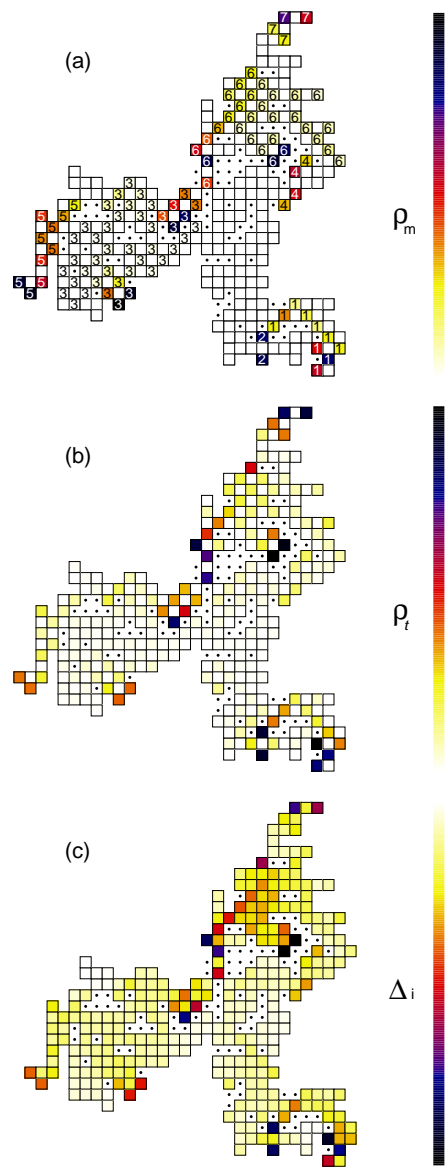}
\caption{(Color online) Properties of a typical cluster with $288$ sites; the classical monomer density (a), the 
triplet density (b), and the local gap estimate $\Delta_i$ (c). All quantities are shown on a color scale ranging 
from the smallest ($0$ in the case of $\rho_t$ and $\rho_m$) to the largest value. The absolute values are irrelevant for the purpose 
of the discussion here. In (a), the numbers inside the squares label the different classical moment regions. The dots indicate empty 
sites.}
\label{3clusters}
\end{center}
\vskip-3mm
\end{figure}

We study the site dependent triplet density $\rho_i=\langle n_t(i)\rangle$, where the triplet occupation number 
$n_t(i)$ is defined such that if a triplet bond connects sites $i$ and $j$, then $n_t(i)=1$ and $n_t(j)=1$, while
$n_t(k)=0$ for all other sites $k$. In addition to visually examining the triplet density for representative individual 
clusters, it is also useful to have a quantitative measure of localization. For this purpose, we use the inverse 
participation ratio (IPR) corresponding to the triplet density;
\begin{equation}
  R_t=\frac{\Big(\sum_{i=1}^{n}\rho_i\Big)^2}{\sum_{i=1}^{n}\rho_i^2} = \frac{4}{\sum_{i=1}^{n}\rho_i^2}.
  \label{iprdefinition}
\end{equation}
This quantity characterizes the number of sites involved in a triplet excitation, and can be averaged over 
cluster realizations. Two extreme cases can help to clarify the meaning of $R_t$; if the triplet is completely 
localized on only two sites, then $R_t=2$, while if it is equally spread out over all the sites of an $n$-site
cluster, then $R_t=n$. We will study the dependence of $\langle R_t\rangle$ on the cluster size.

\begin{figure}
\begin{center}
\includegraphics[width=6.5cm, clip]{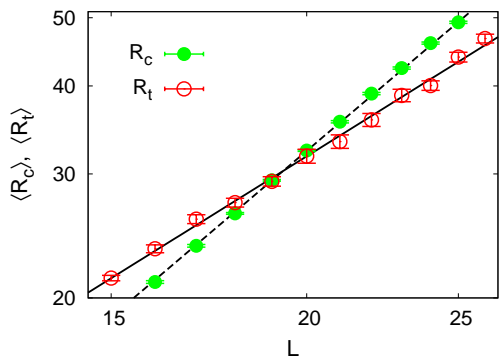}
\caption{Log-log plot showing the finite-size scaling of the IPR of the triplets ($R_t$) and the classical monomers ($R_c$). 
The lines correspond to scaling $L^\gamma$, with $\gamma=1.39(3)$ (based on a line fit) for $R_t$ and $\gamma=D_f$ for $R_c$.}
\label{rtrc}
\end{center}
\vskip-3mm
\end{figure}

In a projector method based on a power $H^P$, one converges to the lowest state in a given symmetry sector when the power 
$P$ of the hamiltonian is sufficiently high. For an $n$-site cluster, one would expect that the $P$ required for convergence 
scales as $n$ or worse. This can be seen if we compare with an alternative projection method---the imaginary-time evolution 
${\rm e}^{-\beta H}|\Psi\rangle$ of the trial state. Here $\beta$ is analogous to an inverse temperature; 
starting from a state at some ``temperature'', we ``cool it'' by increasing $\beta$. 
A better trial state corresponds to a lower initial temperature. For large $\beta$, the 
dominant power $\tilde P$ in a Taylor expansion of ${\rm e}^{-\beta H}$ is $\tilde P=\beta |E_0|$, where $E_0$ 
is the ground state energy, which is proportional to the cluster size $n$. Thus, if we project with just a 
fixed power $P$ of $H$, we would get essentially the same result if $P \approx \tilde P \propto \beta n$. The 
energy scale of the excitations decrease with increasing $n$ (very quickly so in the problem under consideration here,
because the dynamic exponent is large), and we should therefore expect to need larger $\beta$ for larger $n$. 
Thus, in the fixed-power scheme, the $P$ required for convergence should increase as some power (larger
than one) of $n$.

We show examples of the convergence of the IPR for three different clusters in Fig.~\ref{vbsimulation}. The large fluctuations 
in the finite-size gap, discussed in Sec.~\ref{moredata}, naturally also imply large variations in the convergence rate of the 
triplet IPR (which is governed by the gap between the first and second triplet, which also exhibits large fluctuations).
As explained in Sec.~\ref{ipr}, we are restricted to $P$ for which the triplet survival probability in the projection
is reasonably large. In order to ensure that the results truly reflect the lowest excitation for each cluster, we carry out extrapolations 
to infinite $P$ using a simple exponential form, as explained in the caption of Fig.~\ref{vbsimulation}. The fluctuations of the disorder 
averaged $\langle R_t\rangle$ are completely dominated by the cluster-to-cluster variations, and we believe that any remaining 
errors related to the convergence are smaller than the final error bars (based on a few hundred clusters of each size).

\begin{figure}
\begin{center}
\includegraphics[width=7cm]{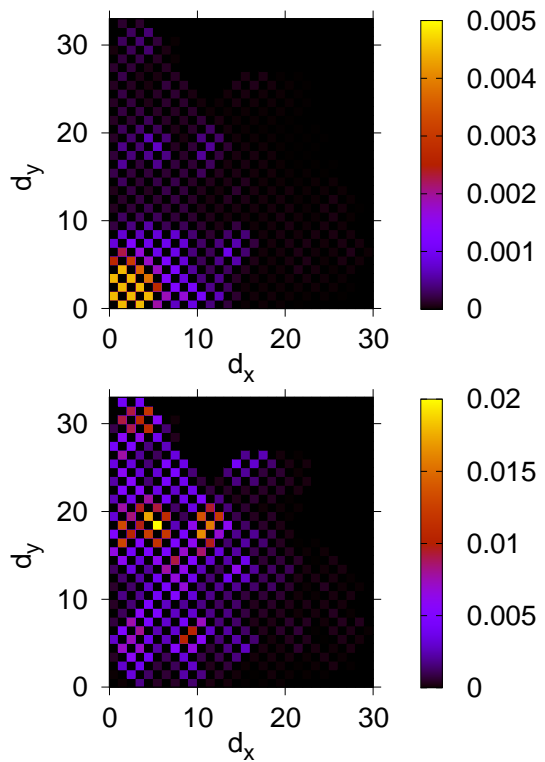}
\caption{(Color online) Probability distribution of the length $(d_x,d_y)$ of the singlet (top panel) and triplet
(bottom panel) bonds for the cluster in Fig.~\ref{3clusters}. The singlet distribution is strongly peaked at
short bonds, and we have therefore cut off the corresponding peak in the low left corner of the histogram. The
the remaining weight is $9\%$ of the total. }
\label{stlengths}
\end{center}
\vskip-3mm
\end{figure}

We first examine the spatial distribution of the triplets. The triplet density for each site of a typical cluster is shown using 
a color scale in Fig.~\ref{3clusters}(b). Here we compare the triplet density with two other calculations---the classical monomer 
density in (a) and the local gap $\Delta_i$ in (c). It is apparent that the triplet is concentrated to a relatively small fraction 
of all the sites of the cluster. At the same time, the affected sites form groups that are spread out over the cluster. This is 
exactly in agreement with our hypothesis of low-energy excitations involving a number of localized moments. 
It is also clear from Fig.~\ref{3clusters} that the classical monomer density is high wherever the triplet probability is 
significant. This proves that our measure of sublattice imbalance in terms of classical monomers indeed corresponds very closely 
to the actual locations affected by excitations. Note also that some sites with high monomer density do not have a high triplet 
density. This is also expected, because the {\it lowest} triplet excitation should not necessarily involve all of the classical 
monomer regions. Higher triplet states may involve other subsets of moments. Finally, there is a very good correspondence 
between regions of low local gaps $\Delta_i$ and hight triplet density. 

Next, we discuss the IPR of the triplet. It is interesting to compare this with the total number of spins in the classical
dimer-monomer model. We therefore also define a classical IPR, as in Eq.~(\ref{iprdefinition}) but with the triplet density 
$\rho_i$ replaced by the classical monomer density. Both these IPRs, averaged over several hundred clusters, are shown versus the 
cluster length $L$ on a log-log scale in Fig.~\ref{rtrc}. They both scale according to power laws. The classical IPR is consistent 
with the form $L^{D_f} \sim \langle n\rangle$. In combination with the fact that the individual moment regions are finite, as we
showed in the previous section, this is in agreement with our extremal-value analysis in Sec.~\ref{extremal}, which relied on the number 
of effective moments being proportional to $n$. However, the triplet IPR scales with a smaller power; $\langle R_t\rangle \sim L^\gamma$
with $\gamma=1.39(3)$. Thus, not all the effective moments are involved in the lowest excitation, but since the size of the 
excitation still grows with $L$ these are not localized excitations.

\begin{figure}
\begin{center}
\includegraphics[width=6.5cm]{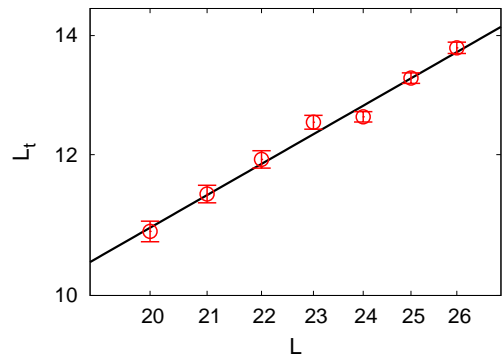}
\caption{(Color online) Size dependence of the average length of the triplet bond on $L \times L$ clusters.
The line corresponds to a power-law divergence $L_t \sim L^\alpha$, with $\alpha=0.81 \pm 0.03$.}
\label{ltscaling}
\end{center}
\vskip-3mm
\end{figure}

Another important aspect of the valence-bond calculation is that the bond lengths also contain information directly pertaining 
to the nature of the excitations. In Fig.~\ref{stlengths} we show the distributions of both the singlet and triplet bonds
lengths for the cluster shown in Fig.~\ref{3clusters}. While the triplet bond is typically long, the singlet distribution is 
strongly peaked for the shortest bonds. We have therefore cut off more than $90\%$ of the weight in the singlet histogram in 
order to be able to show the more interesting distribution of long bonds. Every peak in the triplet distribution can be perfectly
matched to the distance between two regions (on different sublattices) with a high concentration of triplets/monomers in 
Fig.~\ref{3clusters}. This again supports the notion of excitations of weakly interacting effective moments. In the singlet distribution, 
there are also features corresponding to the same lengths as in the triplet case. This is also what one would expect if the triplet state 
essentially corresponds to promoting a long singlet to a triplet in a superposition, as discussed in Sec.~\ref{vbtheory}. The average 
length of the triplet bond also scales with the cluster size according to a power-law, as shown in Fig.~\ref{ltscaling}. This power
law, in combination with that for the triplet IPR (Fig.~\ref{rtrc}) and classical percolation exponents, should be related to the 
dynamic exponent $z\approx 3.6$. Exactly how is not presently clear, however.

\section{\label{perco-summary}Summary and discussion}

To summarize, we have discussed several calculations aimed at elucidating the quantum dynamics of the $S=1/2$ antiferromagnetic Heisenberg 
model on randomly diluted clusters. Quantum Monte Carlo simulations in combination with sum rules show that the low-energy excitations at the 
percolation point are described by an unexpectedly large dynamic exponent; $z \approx 3.6 \pm 0.1 \approx 2D_f$, where $D_f=91/48$ is the 
fractal dimension of 2D percolation. Using extremal-value statistics, we were able to relate $z$ and two exponents characterizing the probability 
distribution of local gaps (an exponent $a$ governing the size dependence and $\omega$ describing the distribution of small gaps), 
according to Eq.~(\ref{zrelation}). This kind of scaling indicates that the excitations involve effective localized finite 
magnetic moments, which interact through the remaining, magnetically inert parts of the cluster. This is also confirmed directly by imaging 
the spatial distribution of the triplet excitations in the valence bond basis, where the triplet state can be described in terms of a pair 
of spins forming a triplet in a ``singlet soup'' of valance bonds. The triplet bond fluctuates between several isolated regions, and its 
average length scales as a power of the cluster size. The average number of spins affected by the excitation also grows as a power $n^{\beta}$ of 
the system size, with $\beta\approx 0.74$. The excitations are thus localized at multiple moment regions, which are spread out over the cluster. 
All these results lead to a picture of an effective low-energy system consisting of a network of globally entangled local moments, where the 
moments correspond to regions of sublattice imbalance.

We have introduced a quantitative measure of sublattice imbalance, in terms of a classical dimer-monomer aggregation model. Monte Carlo
simulations of this model show that the monomers form isolated finite regions, and the number of such regions scales linearly in the cluster size $n$.
Sites with a high triplet concentration coincide very well with high monomer density, confirming directly that sublattice imbalance in the 
Heisenberg model is associated with the formation of weakly interacting effective moments. 

We have also shown that when two identical clusters are coupled in a bilayer, with a small inter-layer coupling $J_\perp$ (smaller than the value 
at which the long-range order vanishes~\cite{andersPRL2002}), the low-energy excitations change dramatically. A finite-size scaling analysis for 
clusters with $J_\perp/J= 0.01$ show a much smaller dynamic exponent, $z \approx D_f$, than for the single-layer clusters. There is no sublattice 
imbalance in this ``dimer  diluted'' bilayer model, and the results therefore provide additional evidence for the important role played by effective 
moments at imbalanced regions in the single-layer clusters.

The result $z\approx 3.6$ was obtained by studying clusters in which the sublattices are balanced globally, i.e., the ground state spin is 
$S=0$. We have also pointed out that for clusters with global imbalance in the sublattice occupation (leading to $S>0$), the dynamic structure factor has 
spectral weight predominantly arising from the $S\to S+1$ channel, while there is very little weight in the $S \to S$ and $S\to S-1$ channels (apart 
from the elastic $S \to S$ weight). Experiments directly probing the inelastic spectral weight, e.g., neutron scattering and nuclear magnetic 
resonance, should be dominated by the $S\to S+1$ channel, and $S$ should be typically large ($\propto \sqrt{n}$) for random clusters. For this 
situation our sum rule method gives a smaller effective dynamic exponent, $z \approx 1.5D_f$, than for the $S=0 \to 1$ excitations of globally 
balanced clusters. The lowest-energy excitations, which are in the $S\to S-1$ channel, are not accessible with the sum rule approach. We have argued, based 
on an analysis of approximate (variational) valence bond states, that their energy should scale with the same $z\approx 3.6\approx 2D_f$ as in the 
case of $S = 0 \to 1$ excitations. 

There have been attempts previously to determine the dynamic exponent of the diluted Heisenberg model based on quantum Monte Carlo calculations. 
Yu {\it et al.} studied the temperature dependence of the correlation length and concluded that it scales in a way consistent with $z=D_f$ at the percolation 
point.\cite{yuPRL2005} However, the scaling assumption was one corresponding to quantum-criticality, which may not apply because the 
percolating cluster at $T=0$ does not have quantum critical fluctuations in the sense of power-law decaying correlation functions. A similar treatment of 
the clean 2D Heisenberg model would fail to give $z=D$ (which is not a quantum-critical exponent but one characteristic of the quantum-rotor excitations 
of the N\'eel state), because the correlation length diverges exponentially as $T \to 0$.\cite{chn88} 

We have here focused exclusively on the dynamics of the percolating cluster. In order to relate the results quantitatively to specific 
experiments, one should include the contributions from all clusters. The cluster distribution is given by classical percolation theory,\cite{stauffer}
which can be combined with the finite-size scaling properties that we have found here for the distributions of the local and smallest gaps. We have also 
not discussed the consequences of our $T=0$ results for the $T>0$ behavior. This is of course also an important experimental issue,
and will be interesting to consider in future studies. 

Moving away from the percolation point $p^*$, our calculations show that the dynamic exponent of the single layer is $z\approx 2=D$. However, the classical 
dimer-monomer model has finite localized monomer regions also away from the percolation point. This suggests that the change in the spin dynamics upon moving 
away from $p^*$ is related to the effective interactions between the moment regions, not the disappearance of the moments. Such a qualitative change in the
interaction aspects of the moments is not completely unexpected, since the spin stiffness of the percolating cluster at $p^*$ is strictly zero in 
the thermodynamic limit \cite{andersPRB2002,moorePRB2004,moorePRB2006} (although the cluster is ordered), whereas it becomes finite (according to a power-law) 
away from the percolation point. The more robust cluster order for $p<p^*$ should qualitatively change the effective interactions between distant monomer 
regions, likely locking all of them to the global N\'eel vector (which is the case for a single moment in a 2D system \cite{sachdev99,hoglund04}). The 
effectively independent nature of the magnetic moments exactly at the percolation point (and the very weak interactions between them, are thus intimately 
related to the fractal structure and related vanishing spin stiffness of the network connecting the moment regions. We presented some results showing 
the cross-over from scaling controlled by the percolation point to 2D behavior. 

In spite of the close agreement with the dynamic exponent expected based on the quantum rotor mechanism \cite{vojtaPRL2005} for the single layer away 
from $p^*$ and the bilayer at $p^*$, it is still not certain that the lowest-energy excitations in these systems are quantum rotor states. Over the years 
there have been considerable efforts to understand the dynamics of various randomly diluted systems close to and at the percolation point. The ``fracton'' 
has been introduced as a generic excitation which develops out of plane waves (e.g., spin waves for an antiferromagnet) for a translationally invariant 
system upon dilution.\cite{orbachModernPhy1994} Numerical calculations based on spinwave theory show that the dynamic exponents for fractons in the 
2D percolating antiferromagnet is very close to $D_f$.\cite{terao94} This calculation does not properly account for the vanishing spin stiffness 
of the percolating cluster at $p^*$ and the existence of localized moments, but it may still be valid close to the percolation point (where a finite stiffness 
develops). It is  possible that the dynamic exponents $z \approx D=2$ and $z \approx D_f$ that we have obtained here for the single layer with $p < p^*$ and 
the bilayer at $p^*$, respectively, are due to fractons, not quantum rotors. However, exactly at the percolation point, the physics of the globally entangled 
local moments that we have discussed here is clearly different from fractons (which can exist in systems that do not have any objects corresponding 
to localized moments) and the value of the dynamic exponent is twice that expected based on fractons.\cite{terao94} It is thus possible that several types of 
excitations co-exist at and close to the percolation point; quantum rotors, fractons, and the globally entangled moment excitations.

One may still be able to describe the low-energy physics of the system away from the percolation point as a network of weakly interacting moments, but now 
in the presence of a staggered field mimicking the coupling to a common N\'eel vector (i.e., the sign of the field depends on which sublattice a moment is 
associated with on the original lattice). The strength of the effective staggered field (which for a finite cluster should be allowed to have a fluctuating
direction as well \cite{hoglund04}) should increase upon moving away from the percolation point (being $0$ at $p^*$, due to the vanishing spin stiffness---the 
energy scale of twisting the N\'eel order globally). Most likely, even an infinitesimal field will asymptotically (for large clusters) change the dynamic exponent.

The role of effectively isolated spins in the formation of long-range order \cite{andersPRB2002} on the percolating cluster has been pointed out by Bray-Ali 
et al.~\cite{moorePRB2004,moorePRB2006} Although some spins can be very weakly coupled (effectively) to the rest of the cluster, correlations between them 
can be stronger than within the backbone of the cluster. Arbitrarily weakly coupled moments formed by groups of spins can also correlate over long distances, 
and hence even a "floppy" fractal cluster (one with vanishing spin-stiffness) can order at $T=0$. This picture seemingly contains some of the 
ingredients of our entangled moments picture. However, the same ordering mechanism was argued to apply both to antiferromagnets and systems of coupled quantum 
rotors, whereas we have shown here that the bilayer (which should correspond more closely the coupled rotor system, since there is no sublattice imbalance) and 
the single layer behave dramatically different. Since the excitations of the single layer cluster are also much lower in energy than the quantum rotor states 
considered in previous discussions of the dynamic exponent,\cite{vojtaPRL2005} a theoretical treatment within a quantum rotor picture is clearly not adequate. 
In Refs.~\onlinecite{moorePRB2006} and \onlinecite{antonioPRB2004} the excitations of the diluted system were analyzed using spin-wave theory, but this method 
also does not capture the significance of almost isolated moments and their long-range global entanglement, and no unusually low energy scale was discussed.

In field-theory language, the ``dangling'' spins, or regions of sublattice imbalance, that we have discussed here correspond to uncompensated Berry 
phases.\cite{sachdevbook} Although it is quite clear that these should exist in diluted quantum antiferromagnets, how to properly take them into account 
in analytical calculations for these systems is not well understood. To our knowledge, the resulting globally entangled moments excitations that we have 
argued for here have not been discussed previously in the literature. The effective low-energy system is similar to the random antiferromagnet considered
by Bhatt and Lee, \cite{bhatt} and also by Sachdev and Ye.\cite{sachdev93} However, there is an important difference in that the nearest-neighbor 
interactions in our system are not frustrated. An effective low-energy hamiltonian should then also not be frustrated. 

The Bhatt-Lee calculation \cite{bhatt} was focused on the thermodynamic properties and did not address the dynamic exponent. The method applied was a 
generalization of the strong disorder renormalization (singlet decimation) scheme by Ma, Dasgupta, and Hu,\cite{ma} which has been applied to numerous 
random antiferromagnetic Heisenberg systems.\cite{fisher,westerberg,Lin} It would be interesting to apply this method also to the diluted clusters. 
However, there is a technical problem in doing this directly, since the decimation scheme is based on random couplings (successively eliminating the strongest 
coupled spin pair and including their remaining effects as modified couplings calculated perturbatively), whereas in the diluted system all couplings are the 
same. It may be possible to carry out a decimation procedure by eliminating strongly {\it correlated} spins, instead of strongly {\it coupled} ones. The 
correlations could be computed perturbatively based on regions of a 
small number of spins, or using quantum Monte Carlo simulations. This way, one could study the renormalization flows of the correlations and how they relate
to the sublattice imbalance that we have quantified here in terms of the classical dimer-monomer systems. The final stages of the decimation procedure should 
lead to bonds (entanglement) between the sites on which our projector QMC calculations give a high triplet probability. However, we have shown that there are 
large fluctuations in the long bonds (singlet as well as triplet) and it is therefore clear that the scheme cannot asymptotically give the correct ground state 
and low-energy excitation in terms of a single bond configurations. In one dimension, the final ``random singlet state'' is known to be asymptotically 
exact,\cite{fisher} in the sense that a single bond configuration is a good representation of a superposition including fluctuations around this reference 
state.\cite{tran09} With large fluctuations of long valence bonds among many moments in the percolating clusters, it seems unlikely that a single reference 
configuration would be a good approximation in this case. It would still be interesting to investigate the flow of the renormalized coupling distribution.

To go further in developing an understanding of the excitations of the weakly interacting effective moments,
instead of working with the full percolating clusters it may be better to explicitly construct the effective low-energy hamiltonian we have discussed here. 
While the geometrical locations of the moments could be obtained using the classical dimer-monomer model, the effective couplings are 
more challenging. One approach would be to just study a bipartite network of spins with some suitable form of the interactions (which should be non-frustrated, 
with antiferromagnetic couplings between sublattices and ferromagnetic intra-sublattice couplings). In principle the spins should have mixed $S$. In one 
dimension such a system is known to have different properties than the random $S=1/2$ chain with only antiferromagnetic couplings.\cite{westerberg} The effective
system could be studied with the methods used here, as well as with the strong-disorder decimation scheme. Comparing results for the moment network with the 
dynamic exponents we have extracted here for the full cluster system (and investigating the robustness of the exponents to variations in the couplings) 
could shed further light on this challenging problem.

\acknowledgments

We would like to thank Yu-cheng Lin and Antonio Castro Neto for useful discussions. This work was supported by the 
NSF under Grant No.~DMR-0803510. Most of the numerical computations were carried out at the Center for Computational 
Science at Boston University.

\end{document}